\documentclass[rd,showpacs,twocolumn]{revtex4-1}
\usepackage{amsfonts,amsmath,amssymb,amsthm}
\usepackage[bookmarks, bookmarksopen, bookmarksnumbered]{hyperref}

\usepackage{graphicx}
\usepackage{color}
\usepackage[utf8]{inputenc}
\usepackage[T1]{fontenc}
\usepackage{lmodern}
\usepackage[right]{rotating}
\usepackage{longtable}
\usepackage{array}
\usepackage{xfrac}

\newcommand{\diff}{\mathrm{d}}
\newcommand{\codename}[1]{\texttt{#1}}

\newcommand{\GRHydro}{\codename{GRHydro}~}

\graphicspath{{./}}

\begin{document}
\title{Stiffness effects on the dynamics of the bar-mode instability of 
       Neutron Stars in full General Relativity}
\date{\today}

\author{Frank \surname{L\"offler}}
\affiliation{Center for Computation \& Technology, Louisiana State University, Baton Rouge, LA 70803 USA}
\author{Roberto \surname{De Pietri}}
\affiliation{Parma University and INFN Parma, via G.P. Usberti 7/A, I-43124 Parma (PR), Italy}
\author{Alessandra \surname{Feo}}
\affiliation{Parma University and INFN Parma, via G.P. Usberti 7/A, I-43124 Parma (PR), Italy}
\author{Luca \surname{Franci}}
\affiliation{Dipartimento di Fisica e Astronomia, Universit\'a di Firenze - INFN Sezione di Firenze,
I-50019 Sesto Fiorentino (FI), Italy}
\author{Francesco  \surname{Maione}}
\affiliation{Parma University and INFN Parma, via G.P. Usberti 7/A, I-43124 Parma (PR), Italy}

\begin{abstract}
We present results on the effect of the stiffness of the equation of state on the
dynamical bar-mode instability in rapidly rotating polytropic models of neutron
stars in full General Relativity. We determine the change in the threshold for
the emergence of the instability for a range of the adiabatic $\Gamma$ index from
2.0 to 3.0, including two values chosen to mimic more realistic equations of
state at high densities.
\end{abstract}

\LTcapwidth=\columnwidth

\pacs{
04.25.D-,  
04.40.Dg,  
95.30.Lz,  
97.60.Jd   
}

\maketitle

\section{Introduction}
\label{sec:intro}
Non-axisymmetric deformations of rapidly rotating self-gravitating
objects are a generic phenomenon in nature and are expected to appear in a
wide range of astrophysical scenarios, like stellar core
collapses~\cite{Shibata:2004kb,Ott:2005gj}, accretion-induced collapses
of white dwarfs~\cite{Burrows:2007yx}, or mergers of two neutron
stars~\cite{Shibata:2003ga,Shibata:2005ss}. Over more than a decade, a
considerable amount of work has been devoted to the search of unstable
deformations that, even when starting from an axisymmetric configuration, can
lead to a highly deformed, rapidly rotating, massive
object~\cite{Shibata:2000jt,Baiotti:2006wn,Kruger:2009nw,
Kastaun:2010vw,Lai:1994ke,DePietri:2014mea}.
In the case of neutron stars, such deformations would lead to an intense
emission of gravitational waves in the kHz range, potentially detectable on
Earth within the next decade~\cite{LIGOVIRGO:2013} by next-generation
gravitational-wave detectors such as Advanced LIGO~\cite{Harry:2010}, Advanced
VIRGO, or KAGRA~\cite{KAGRA:2012}.

Any insight on the possible astrophysical scenarios where such instabilities might
be present would aid potential observations and their analysis and understanding.
It is well-known that rotating neutron stars are subject to
non-axisymmetric instabilities for non-radial axial modes with
azimuthal dependence $\mathrm{e}^{i m \phi}$ (with $m = 1,2,\ldots$)
when the instability parameter $\beta \equiv T/|W|$ (i.e. the ratio
between the kinetic rotational energy $T$ and the gravitational
potential energy $W$) exceeds a certain critical value $\beta_c$. This
instability parameter plays an important role in the study of the
so-called dynamical bar-mode instability, i.e. the $m=2$ instability
which takes place when $\beta$ is larger than the threshold $\beta_c$
~\cite{Baiotti:2006wn}. Previous results for the onset of
the classical bar-mode instability have already shown that the
critical value $\beta_c$ for the onset of the instability is not a
universal quantity and it is strongly influenced by the rotational
profile~\cite{Shibata:2003yj,KarinoEriguchi03}, by relativistic
effects~\cite{Shibata:2000jt,Baiotti:2006wn}, and, in a quantitative way, by the
compactness~\cite{Manca:2007ca}.

\begin{figure}
\begin{centering}
  \includegraphics[width=0.45\textwidth]{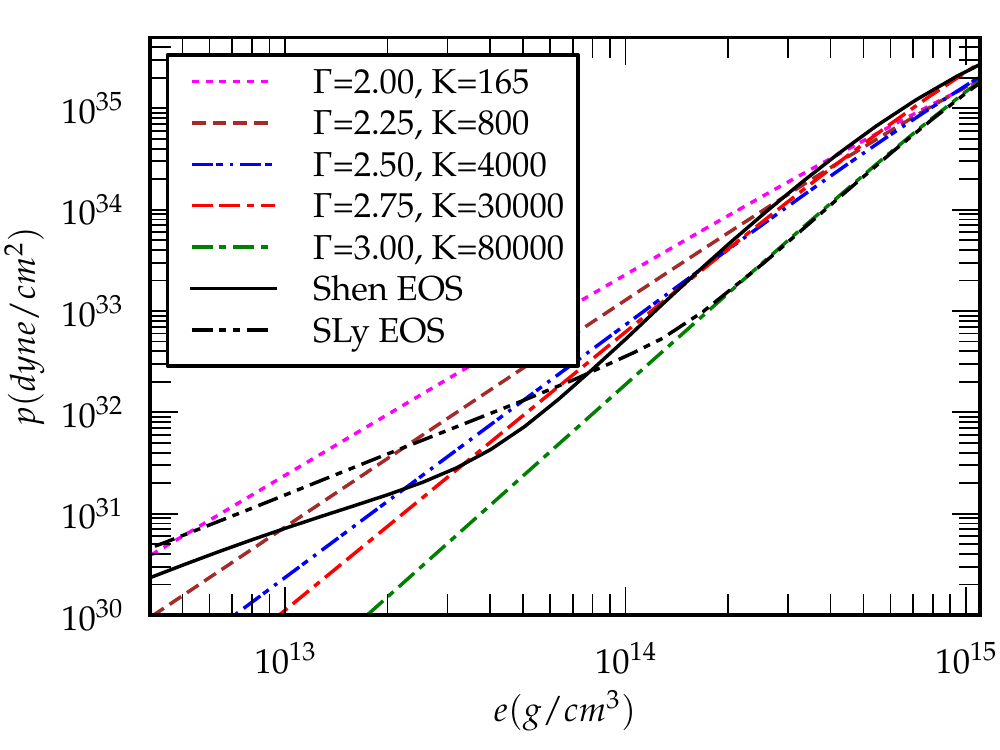}\\
\end{centering}
\caption{Diagram of the pressure $P$ vs. the energy density $e$ for all polytropic
         EOSs considered and two {\it realistic} EOSs for nuclear matter, namely
         (1) the Shen proposal~\cite{shen98,shen98b}; (2) the unified SLy
         prescription~\cite{Douchin:2001sv}. The polytropic
         EOS with $\Gamma=2.75$ is close to the Shen EOS, while the case of
         $\Gamma=3.0$ is close to the SLy EOS, both above a certain density.}
\label{fig:EOSs}
\end{figure}

However, until the recent work in~\cite{DePietri:2014mea}, significant
evidence of their presence when realistic EOSs are considered was missing.
For example, in~\cite{Corvino:2010},
using the unified SLy EOS~\cite{Douchin01}, the presence of a shear-instability
was shown, but there was no sign of the classical bar-mode instability and its
critical behavior.
The aim of the present work is to get more insight
into the behavior of the classical bar-mode instability when the matter is
described by EOSs with different stiffness. The
investigations in the literature into its dependence on the stiffness of EOSs usually focused 
on values of $\Gamma$
(i.e. the adiabatic index of a polytropic EOS) in the range from
$1$ to $2$~\cite{Lai:1994ke,2007PhRvD..76b4019Z,Kastaun:2010vw}, 
while the expected value for a real neutron
star is more likely higher, between $2$ and $3$, and probably
around $\Gamma=2.75$ (at least in large portions
of the interior). Such a choice for the EOS was already
implemented in~\cite{Oechslin2007aa}, and also quite
recently in~\cite{Giacomazzo:2013uua,DePietri:2014mea}.
Its benefit is the ability to maintain the simplicity
of a polytropic EOS and yet to obtain properties that resemble a more
realistic case. Indeed, as it is shown in Fig.~\ref{fig:EOSs}, a
polytropic EOS with $K=30000$ and $\Gamma=2.75$ is qualitatively
similar to the Shen proposal~\cite{shen98,shen98b} in the density
interval between $2 \times 10^{13} \text{g/cm}^3$ and $10^{15}
\text{g/cm}^3$, while a polytropic EOS with $K=80000$ and $\Gamma=3.00$
approximately resembles the SLy EOS for densities higher than
$2\times10^{14} \text{g/cm}^3$.

The organization of this paper is as follows. In Sec.~\ref{sec:setup}
we describe the properties of the relativistic stellar models we
investigated and briefly review the numerical setup used for their
evolutions. In Sec.~\ref{sec:results} we present and discuss our
results, showing the features of the evolution and
quantifying the effects of the compactness on the onset of the
instability. Conclusions are given in
Sec.~\ref{sec:conclusions}. Throughout this paper we use a
space-like signature $-,+,+,+$, with Greek indices running from 0 to
3, Latin indices from 1 to 3, and the standard convention for summation
over repeated indices. Unless otherwise stated, all quantities are
expressed in units in which $c=G=M_\odot=1$.

\section{Initial models and Numerical setup}
\label{sec:setup}

We follow the same setup as in~\cite{DePietri:2014mea} (only changing the
EOS parameters $\Gamma$ and $K$), but for convenience the main ideas are
summarized in the following section.

In this work we solve Einstein's field equations
\begin{equation}
 G_{\mu\nu} = 8\pi T_{\mu\nu} \, , 
\end{equation}
where $G_{\mu\nu}$ is the Einstein tensor of the four-dimensional
metric $g_{\mu\nu}$ and $T^{\mu\nu}$ is the stress-energy 
tensor of an ideal fluid. The energy-momentum tensor $T^{\mu\nu}$ can be parametrized as 
\begin{equation}
 T^{\mu\nu} = \rho\left(1+\epsilon+\frac{P}{\rho}\right)u^\mu u^\nu + Pg^{\mu\nu} \, ,
\end{equation}
where $\rho$ is the rest-mass density, $\epsilon$ is the specific internal energy of the matter,
$P$ is the pressure, and $u^\mu$ is the matter $4$-velocity.
The evolution equations for the matter follow from the conservation laws for
the energy-momentum tensor $\nabla_\mu T^{\mu\nu}=0$ and the baryon number 
$\nabla_\mu(\rho u^\mu)=0$, closed by an EOS of the type $P=P(\rho,\epsilon)$.

In order to generate the initial data, we use a
$\Gamma$-type EOS of the form
\begin{equation}
 P = K\rho^\Gamma \, ,
 \label{eq:eosP_Gamma}
\end{equation} 
where the following relation between $\epsilon$ and $\rho$ holds:
$\epsilon=K \rho^{(\Gamma-1)} / (\Gamma-1)$. On the other hand, the
evolution is performed using the so-called {\it ideal-fluid}
($\Gamma$-law) EOS
\begin{equation}
 P = (\Gamma-1) \rho \epsilon \, ,
 \label{eq:eosP_Ideal}
\end{equation} 
that allows for increase of the internal energy by shock heating, if shocks are present.

We solve the above set of equations using the usual $3+1$
space-time decomposition, where the space-time is foliated as a tensor
product of a three-space and a time coordinate $t$ (which is selected
to be the $x^0$ coordinate). In this coordinate system the metric can
be split as $g^{\mu\nu}=-n^\mu n^\nu + h^{\mu\nu}$, where $h^{\mu\nu}$
has only the spatial components different from zero and can be used to
define a Riemannian metric $\gamma^{ij}=h^{ij}$ on each foliation. The
vector $n^\mu$, that determines the direction normal to the
3-hypersurfaces of the foliation, is decomposed in terms of the lapse
function $\alpha$ and the shift vector $\beta^i$, such that
$n^\mu=\alpha^{-1}\cdot(1,\beta^i)$.  We also define the fluid
three-velocity $v^i$ as the velocity measured by a local zero-angular
momentum observer ($u^i=\alpha v^i - \beta^i$), while the Lorentz
factor is $\alpha u^0=\sqrt{1-\gamma_{ij}v^iv^j}$. Within this
formalism, the conservation of the baryon number suggests the use of
the conserved variable $D=\sqrt{\gamma}\alpha u^0 \rho$ with the
property that $\int D \, d^3x = \text{const}$ along the
time-evolution $t$.

\subsection{Initial Data}
\label{sec:models}

The initial data of our simulations are calculated as stationary
equilibrium solutions for axisymmetric and rapidly rotating
relativistic stars in polar coordinates~\cite{Stergioulas95}.
We assume that the metric
describing the axisymmetric and stationary relativistic star has the
form
\begin{align}
 ds^2 = -e^{\mu + \nu} dt^2 + e^{\mu - \nu} r^2 & \sin^2\theta(d\phi -
 \omega dt^2)^2 \nonumber \\ & + e^{2 \xi} (dr^2 + r^2 d \theta^2) \, ,
\end{align}
where $\mu$, $\nu$, $\omega$, and $\xi$ are space-dependent metric
functions.  Similarly, we assume the matter to be characterized by a
non-uniform angular velocity distribution of the form
\begin{equation}
 \Omega_c - \Omega = \frac{1}{ \hat{A}^2 r_e^2 } \bigg[ \frac{ (\Omega
     - \omega) r^2 \sin^2\theta e^{-2 \nu} }{ 1 - (\Omega - \omega)^2
     r^2 \sin^2\theta e^{-2 \nu} } \bigg] \, ,
\end{equation}
where $r_e$ is the equatorial stellar coordinate radius, $\Omega_c$ is the
angular velocity at the center of the star, and the
coefficient $\hat{A}$ is the measure of the degree of the differential
rotation, which we set to be $\hat{A} = 1$, analogous to
works in the literature, and especially ~\cite{DePietri:2014mea}.
Once imported onto the Cartesian grid,
throughout the evolution we compute the coordinate angular velocity
$\Omega$
on the $(x,y)$ plane as
\begin{equation} 
\Omega = \frac{u^\phi}{u^0} = \frac{ u^y \cos\phi - u^x \sin\phi }{ u^0
  \sqrt{ x^2 + y^2 } } \, .
\end{equation}
Other characteristic quantities of the system such as the baryon mass
$M_0$, the gravitational mass $M$, the internal energy
$E_{\textrm{int}}$, the angular momentum $J$, the rotational kinetic
energy $T$, the gravitational binding energy $W$ and the instability
parameter $\beta$ are defined as~\cite{Baiotti:2006wn}:
\begin{align}
 M_0 & \equiv \int d^3x D \, , \\ M & \equiv \int d^3x (-2 T^0_0 +
 T^\mu_\mu ) \alpha \sqrt{\gamma} \, , \\ E_{\textrm{int}} & \equiv
 \int d^3x D \varepsilon \, , \\ J & \equiv \int d^3x \, T^0_{\phi}
 \alpha \sqrt{\gamma} \, , \\ T & \equiv \int d^3x \Omega T^0_{\phi}
 \alpha \sqrt{\gamma} \, , \\ W & \equiv T + E_{int} + M_0 - M \, , \\[1mm]
   \beta & \equiv T/|W| \, ,
\end{align}
where $\alpha \sqrt{\gamma}$ is the square root of the
four-dimensional metric determinant.  Notice that the definitions of
quantities such as $J$, $T$, $W$ and $\beta$ are meaningful only
in the case of stationary axisymmetric configurations and should
therefore be treated with care once the rotational symmetry is lost.
All the equilibrium models considered here have been calculated using
the relativistic polytropic EOS given in Eq.~(\ref{eq:eosP_Gamma}),
and we have chosen the polytropic EOS parameters to be
$\Gamma=[2.0$, $2.25$, $2.5$, $2.75$, and $3.0]$ for the adiabatic index, and
$K=[165$, $800$, $4000$, $30000$, and $80000]$
for the polytropic constant respectively.
This choice allows, for each pair of $\Gamma$ and $K$, a maximum neutron
star mass of $\simeq 2.1 M_\odot$. The exception is $\Gamma=2.75$, where
simulation data from~\cite{DePietri:2014mea}, which allows for higher masses,
were re-used to save computation time, given their identical numerical setup.
Note that these values are different from the ones used in~\cite{Manca:2007ca}
($\Gamma=2.0, K=100$). Note, however, that the choice of $K$ does not change
the results presented in this work. The actual value of the polytropic constant $K$
fixes the overall scale of the physical system; i.e. the assertion that we are generating
and simulating a model with a given baryonic mass $M_0$ is related to the
value chosen for $K$. Indeed, in order to claim that the threshold for the instability
depends on the stiffness of the EOS, we need to eliminate the dependencies on the
dimensional scales as well as on the chosen value of the polytropic constant $K$.
An efficient way to do so is to extrapolate the result for $M_0 \rightarrow 0$, which
corresponds to the Newtonian limit, where the general relativistic
effects can be neglected. Using the same procedure followed
in~\cite{Manca:2007ca}, we choose sequences of constant rest-mass
density models among the following possible values for the total Baryon mass $M_0$ 
($0.5$, $1.0$, $1.5$, $2.0$, and $2.5$ $M_\odot$).
We restrict the values of the instability parameter $\beta$ to the range
$[0.255, 0.272]$, and we leave the analysis of models with
lower values to future work.
The initial conditions for the evolution have been generated using
Nicholas Stergioulas' RNS code~\cite{Stergioulas95}. Any 
model can be uniquely determined by three parameters (once the value of the
differential rotation parameter has been fixed to $\hat{A}=1$). We have decided to 
denote each of the generated models using
the values of the adiabatic index $\Gamma$, the conserved baryonic
mass $M_0$, and the $\beta$
parameter at $t=0$. As a consequence of this choice, in the rest 
of this paper we will refer to a particular model using the
following notation. For example, G2.00M1.5b0.270 will denote a model with an
adiabatic index of $2.0$, a conserved baryonic mass $M_0 = 1.5M_\odot$ and a value of the
initial instability parameter $\beta=0.270$.

One of the main features of the
generated models is that, due to the high rotation, none of them
have the density maximum at the center of the star, but rather
at some distance from it. This means that all
of the models studied are characterized by a {\it toroidal}
configuration, i.e. the maximum of the density is not on the
rotational axis. As has been shown in~\cite{DePietri:2014mea}, there is not always a
correlation between having a toroidal configuration and being unstable
against the dynamical bar-mode instability.

\begin{figure*}
  \vspace{-4mm}
  \begin{centering}
    \begin{tabular}{cc}
      \includegraphics[width=0.49\textwidth, trim=0cm 0cm 0cm 1cm, clip=true]{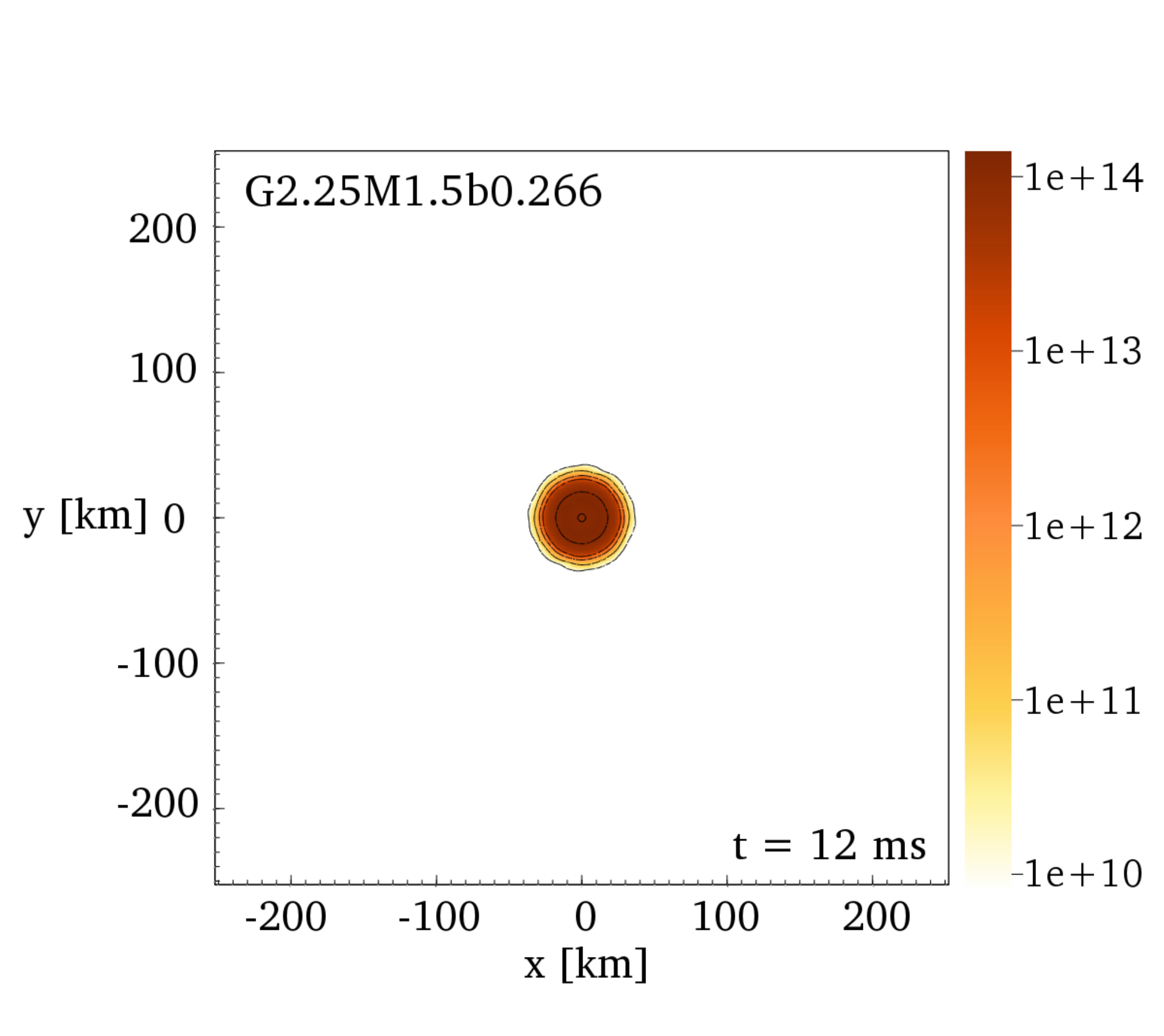}
      \includegraphics[width=0.49\textwidth, trim=0cm 0cm 0cm 1cm, clip=true]{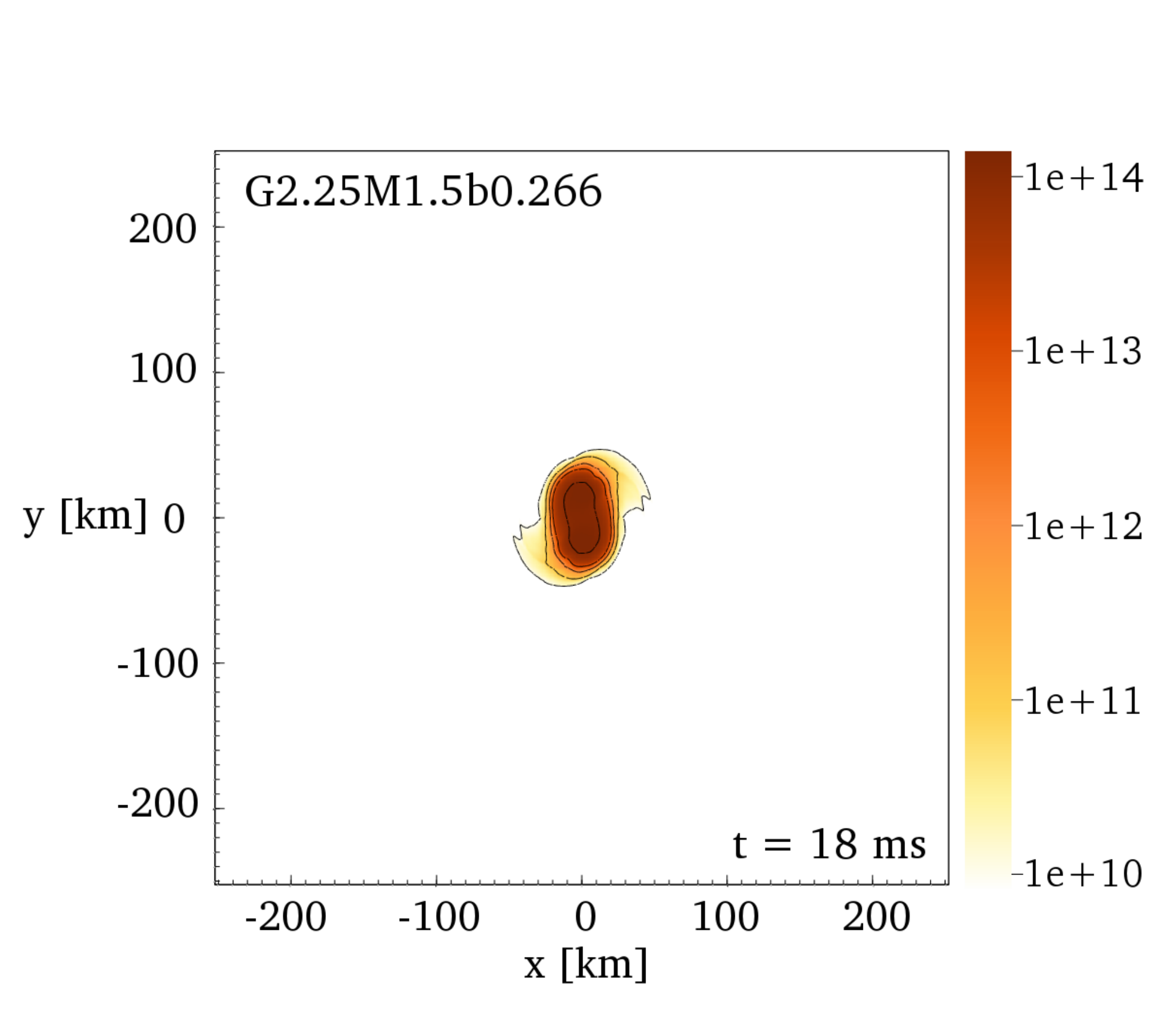}
    \end{tabular}
    \begin{tabular}{cc}
      \includegraphics[width=0.49\textwidth, trim=0cm 0.5cm 0cm 2.5cm, clip=true]{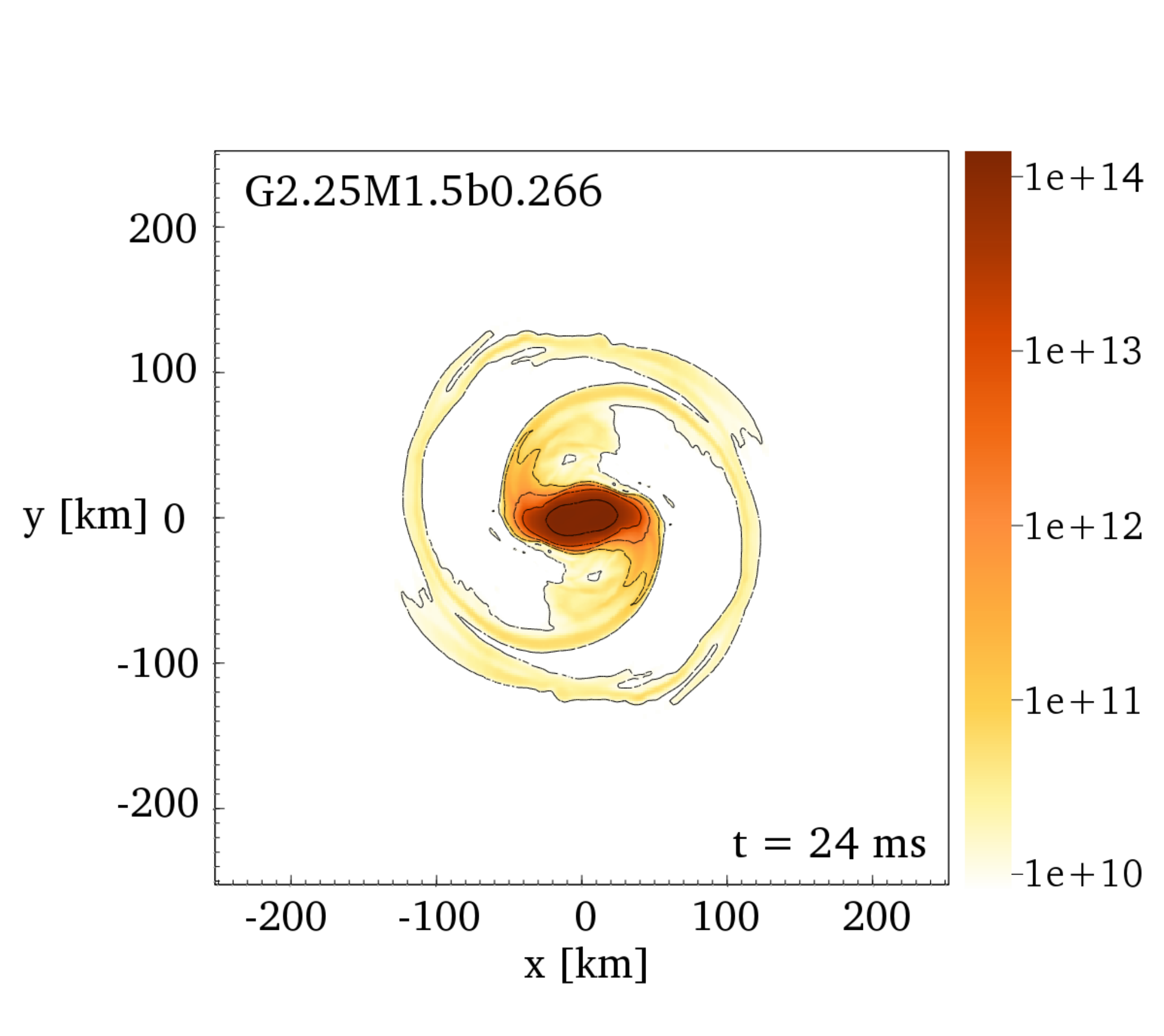}
      \includegraphics[width=0.49\textwidth, trim=0cm 0.5cm 0cm 2.5cm, clip=true]{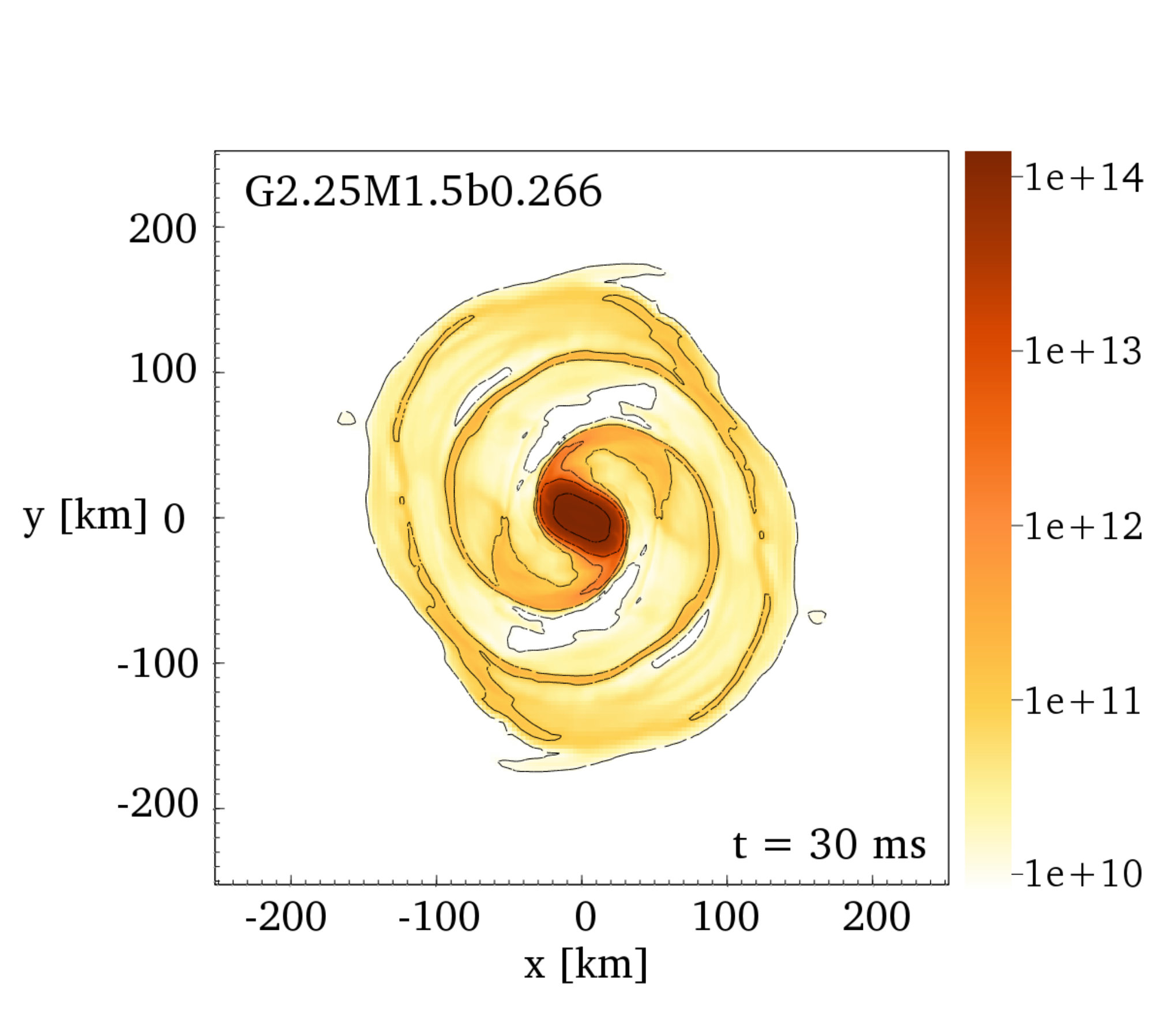}
    \end{tabular}
\end{centering}
\caption{Snapshots of the rest-mass density $\rho$ in the $(x,y)$
  plane for model G2.25M1.5b0.266 at different stages of the
  evolution, namely $t$ = 12 and 18 ms (top row), $t$ = 24 and 30 ms
  (bottom row).  The color code is defined in terms of g/cm$^{3}$.
  Additionally, isodensity contours are shown for $\rho = 10^{10}$,
  $10^{11}$, $10^{12}$, $10^{13}$ and $10^{14}$ g/cm$^{3}$.
\label{fig:snap_shot}}
\end{figure*}

\begin{figure*}
\vspace{-4mm}
\begin{center}
\includegraphics[width=0.98\textwidth]{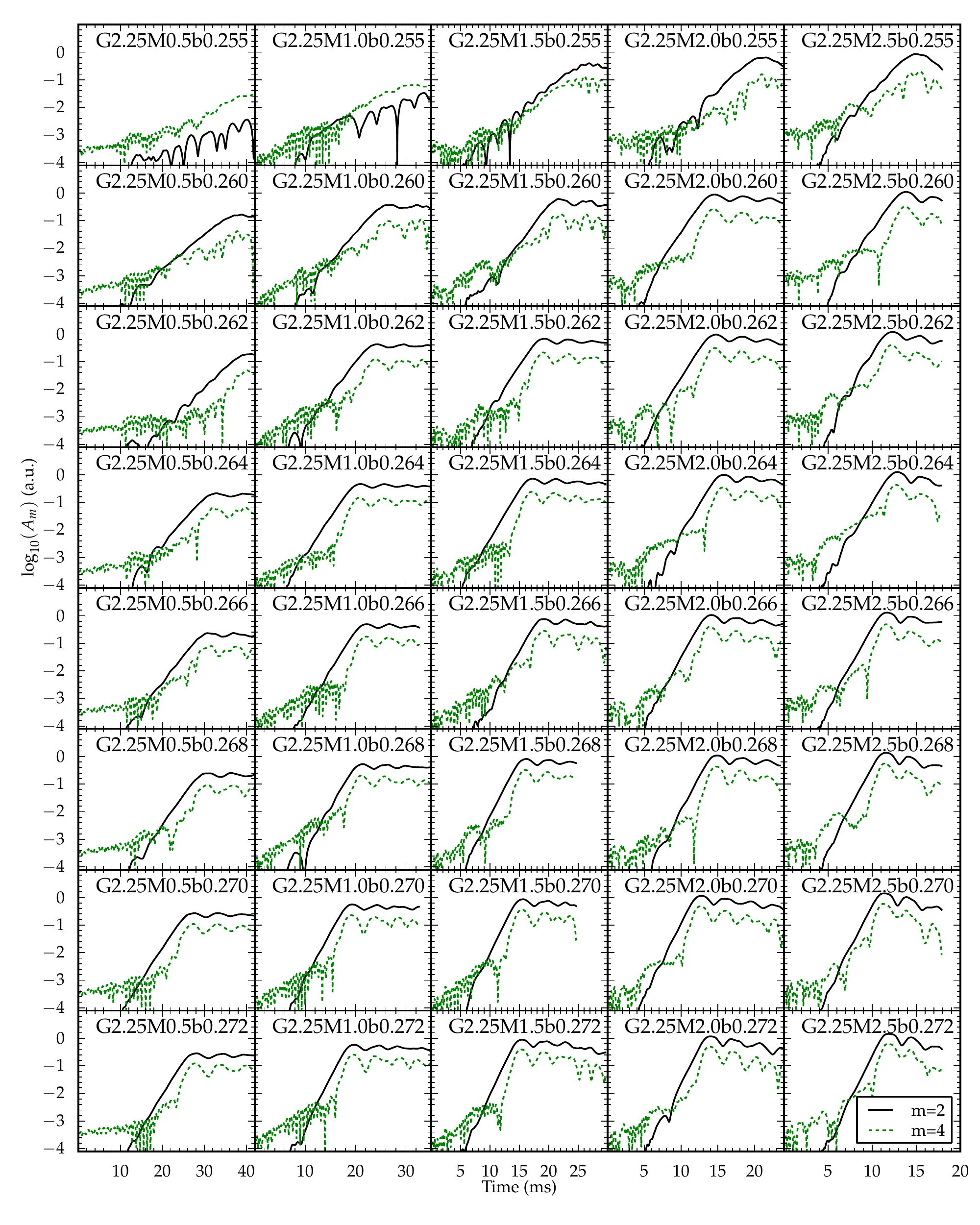}
\end{center}
\vspace{-8mm}
 \caption{Mode dynamics for selected models that are
   characterized by a value of the instability parameter $\beta$
   between 0.255 and 0.272. Almost all models with $\beta \ge 0.255$
   show the typical dynamics one would expect for the dynamical $m=2$ bar
   mode instability. The exceptions are models G2.25M0.5b0.255 and G2.25M1.0b0.255,
   which are not well above $\beta_c$.
\label{fig:high_beta}}
\end{figure*}

\subsection{Numerical setup and evolution method}
\label{sec:numsetup}

We use exactly the same numerical setup as in~\cite{DePietri:2014mea}. Because of this,
we only briefly describe the specific methods used for this
work together with the chosen, relevant, parameters. The reader is
referred to~\cite{Loffler:2011ay} for a description of the Einstein
Toolkit, and to~\cite{DePietri:2014mea} for details about our particular
setup.

The core of the code used for this work is the Einstein
Toolkit~\cite{Loffler:2011ay,EinsteinToolkit:web}, which is a free,
publicly available, community-driven general relativistic (GR) code,
capable of performing numerical relativity simulations that include
realistic physical treatments of matter, electromagnetic
fields~\cite{Moesta:2013dna}, and gravity.

The Einstein Toolkit is built upon several open-source components that
are widely used throughout the numerical relativity community.
Only the ones which were actually used in this work are mentioned below.
Many components of the Einstein Toolkit use the \codename{Cactus}
Computational
Toolkit~\cite{Cactuscode:web,Goodale:2002a,CactusUsersGuide:web}, a
software framework for high-performance computing (HPC\@).

Within this study, the adaptive mesh refinement (AMR) methods
implemented by \codename{Carpet}~\cite{Schnetter:2003rb,
  Schnetter:2006pg, CarpetCode:web} have been used.
Hydrodynamic evolution techniques are provided
by the \GRHydro package~\cite{Baiotti:2004wn,Hawke:2005zw}.

The evolution of the spacetime metric in the Einstein Toolkit is
handled by the \codename{McLachlan}
package~\cite{McLachlan:web}.  This code is
auto-generated by Mathematica using
\codename{Kranc}~\cite{Husa:2004ip,Lechner:2004cs,Kranc:web},
implementing the Einstein equations via a $3+1-$dimensional split
using the BSSN
formalism~\cite{Nakamura:1987zz,Shibata:1995we,Baumgarte:1998te,
  Alcubierre:2000xu,Alcubierre:2002kk}.

Within this paper a fourth-order
Runge-Kutta~\cite{Runge:1895aa,Kutta:1901aa} method was used, and
Kreiss-Oliger dissipation was applied to the curvature evolution
quantities in order to damp high-frequency noise.

We use fourth-order finite difference stencils for the curvature evolution,
$1+\log$~\cite{Alcubierre:2002kk} slicing,
and a $\Gamma$-driver shift condition~\cite{Alcubierre:2002kk}.
During time evolution, a
Sommerfeld-type radiative boundary condition is applied to all
components of the evolved BSSN variables as described
in~\cite{Alcubierre:2000xu}.

All presented results use the Marquina Riemann
solver~\cite{Donat:1996cs,Aloy:1999ne} and PPM (the piecewise
parabolic reconstruction method)~\cite{Colella:1982ee}.  An artificial
low-density atmosphere with $\rho_{\text{atm}}=10^{-10}$ is used, with
a threshold of $\rho_{\text{atm\_reset}}=10^{-7}$ below which regions
are set to be atmosphere.  Hydrodynamical quantities are also set to be
atmosphere at the outer boundary.

All evolutions presented use a mirror symmetry across the $(x,y)$
plane, consistent with the symmetry of the problem, which reduces the
computational cost by a factor of $2$.
Since we are not
interested in investigating whether odd modes play any role, we present 
only results obtained by imposing an additional $\pi$-symmetry, reducing
the computational cost by another factor of $2$.

\section{Results}
\label{sec:results}

As discussed in Sec.~\ref{sec:intro} and~\ref{sec:setup}, the
goal of the present work is to study the matter instability that may develop
in the case of rapidly differentially rotating relativistic star models,
using different configurations of EOSs.
The other important requirement we need to fulfill is that our study has to be 
computationally feasible. To achieve this goal, we need to evolve the largest 
number of models using the available amount of computational resources 
in the most efficient way. 
In selecting a numerical setting we can play with many 
parameters, namely: the location of the outer boundary, the number 
of refinement levels, the size and resolution of the finest grid
and the symmetries to be imposed on the dynamics. 
All the simulations in the present work are performed 
using the same setting for the computational domain. More precisely, we use 
the same setup as in~\cite{DePietri:2014mea}:
three box-in-box (covering the quarter space with $x\ge0$ and $z\ge0$) refinement levels, 
with boundaries at distances of $L=42,84,168$ from the 
origin of the coordinate system and grid spacings $dx,2\,dx,4\,dx$, respectively,
where we set $dx=0.5$ (that correspond to a resolution $dx\simeq 0.738$ km).
Using mirror symmetry across the $x-y$ plane and $\pi$ symmetry across the
$y-z$ plane, this corresponds to a hierarchy of 
three computational grids, each one of size $85\times169\times85$ 
points plus ghost and buffer zones.

We have chosen to use this domain (conservative though large enough to capture the whole global 
dynamics of a bar-mode instability) in order to exclude any influence of the computational setup 
on observed differences between models. The actual size of the finest grid and the 
computational setup is determined by the most demanding models.
Fig.~\ref{fig:snap_shot} shows a few snapshots for the 
evolution of the rest-mass density $\rho$ at different times for
a representative model, namely G2.25M1.5b0.266 which is characterized by
$\Gamma=2.25$, $\beta=0.266$ and $M_0=1.5 \,M_\odot$. This is indeed the typical
evolution one would expect for a stellar model which is unstable against 
the dynamical bar-mode instability.

\subsection{Analysis Methods}
\label{sec:analysis}

In order to compute the growth time of the instability, $\tau_2$, we use the
quadrupole moments of the matter distribution $Q^{ij}$, computed in terms of 
the conserved density $D$ as
\begin{equation}
\label{eq:defQxy}
Q^{ij} = \int\! d^{3}\!x \; D \; x^{i} x^{j} \ .
\end{equation}
In particular, we perform a nonlinear least-square fit of $Q^{xy}$
 (the star spin axis is aligned in the $z$-direction), using the trial function
\begin{equation}
 \label{eq:Qxy}
 Q^{xy}(t) = Q^{xy}_0\mathrm{e}^{\frac{t}{\tau_2}}\cos(2\pi f_2 t+\phi_0) \, .
\end{equation}
Using this trial function, we can extract the growth time $\tau_2$ and the frequency
$f_2$ for the unstable $m=2$ modes.
We also define the modulus $Q(t)$ as
\begin{equation}
 \label{eq:Q}
 Q \equiv\frac{1}{2}\sqrt{(2Q^{xy})^2 + (Q^{xx}-Q^{yy})^2} \, ,
\end{equation}
and the distortion parameter $\eta(t)$ as
\begin{equation}
 \label{eq:def_eta}
 \eta \equiv \frac{2 Q}{(Q^{xx}+Q^{yy})} \, .
\end{equation}
Finally, we decompose the rest-mass density into its spatial
rotating modes $P_m(t)$
\begin{equation}
 P_m \equiv \int\diff^3x \, \rho \mathrm{e}^{\mathrm{i}m\phi} 
\end{equation}
and the ``amplitude'' and ``phase'' of the $m$-th mode are defined as
\begin{equation}
 \label{eq:Am}
A_m = |P_m| \,\quad\, \text{and} \,\quad\, \phi_m \equiv arg(P_m) \, .
\end{equation}
Despite their name, the amplitudes defined in Eq.~(\ref{eq:Am}) do not
correspond to proper oscillation eigenmodes of the star but to global
characteristics that are selected in terms of their spatial azimuthal shape.  
Eqs.~(\ref{eq:Qxy})-(\ref{eq:Am}) are expressed in terms of the
coordinate time $t$, and therefore they are not gauge-invariant. However,
the length scale of variation of the lapse function at any given time
is always small when compared to the stellar radius, ensuring that events
close in coordinate time are also close in proper time.

\subsection{General features of the evolution above the
threshold for the onset of the bar-mode instability}
\label{subsec:above}

The general features of the evolution are common to
all the models that show the expected dynamics in presence of the
bar-mode $m=2$ instability. In Fig.~\ref{fig:high_beta} the
``mode-dynamics'' of most of the studied models with $\Gamma=2.25$ are
shown as an example. For all these models (except for G2.25M0.5b0.255 and G2.25M1.0b0.255)
it is indeed possible to extract the main features of the $m=2$ mode
using the trial function detailed in Eq.~(\ref{eq:Qxy}).
As in~\cite{DePietri:2014mea}, we decided to quantify the properties of
the bar-mode instability by means of a non-linear fit, using
the trial dependence of Eq.~(\ref{eq:Qxy}) on a time interval where the
distortion parameter $\eta$ defined in Eq.~(\ref{eq:def_eta}) is
between $1\%$ and $30\%$ of its maximum value.

The results of all these fits are collected in
Tab.~\ref{tab:models} in the Appendix, where we report for each model the 
maximum value assumed by the distortion parameter $\max(\eta)$, the time
interval $[t_i,t_f]$ selected for the fit, the value
$\beta(t_i)$ corresponding to the value of the instability parameter
$\beta$ at the beginning of the fit interval and $\tau_2$ and $f_2$,
the growth time and frequency that characterize the $m=2$ bar-mode
instability, respectively.

\subsection{Effects of the compactness on the threshold for the onset of the bar-mode 
instability}
\label{subsec:BAR}

We have chosen to investigate the effect of the compactness on the
classical bar-mode instability, following the same
procedure as in~\cite{Manca:2007ca,DePietri:2014mea}, but now for five
stiffness values. We determined the critical value
of the instability parameter $\beta$ for the onset of the instability
by simulating, for each value of the stiffness,
five sequences of initial models having the same value of
$M_0$ but different values of $\beta$. For these simulations we
decided to employ the same resolution $dx=0.5$ on the finest grid 
for all cases. This choice was motivated by the need 
to limit the computational cost.

We now restrict our analysis to the models for which we observed the maximum
value of the distortion parameter $\eta$ to be greater than $0.20$
(see Tab.~\ref{tab:models}).
For these models, we explicitly checked that the reported unstable
modes correspond to the classical bar-mode instability and not to a
shear-instability by ensuring that the frequency of the mode divided by
two is at most only marginally inside of the co-rotation band of the model.

\begin{figure*}
\vspace{-2mm}
\begin{centering}
\includegraphics[width=0.45\textwidth]{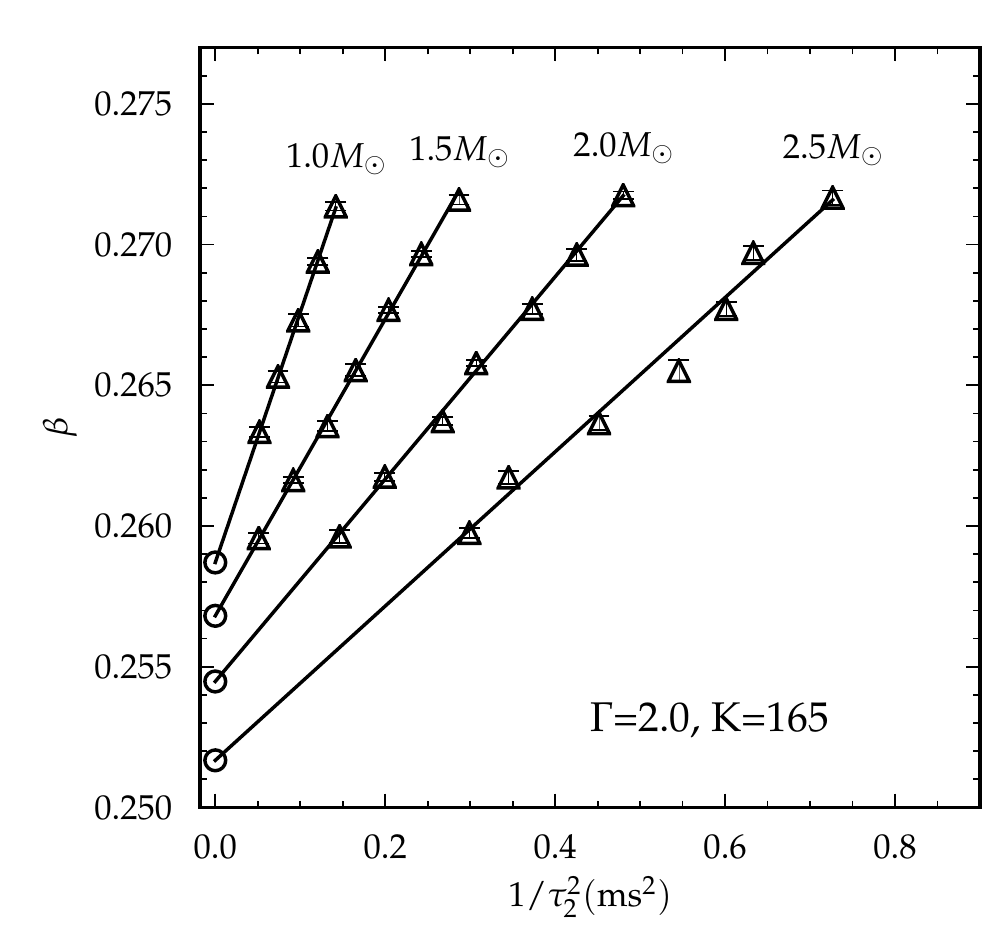}
\includegraphics[width=0.45\textwidth]{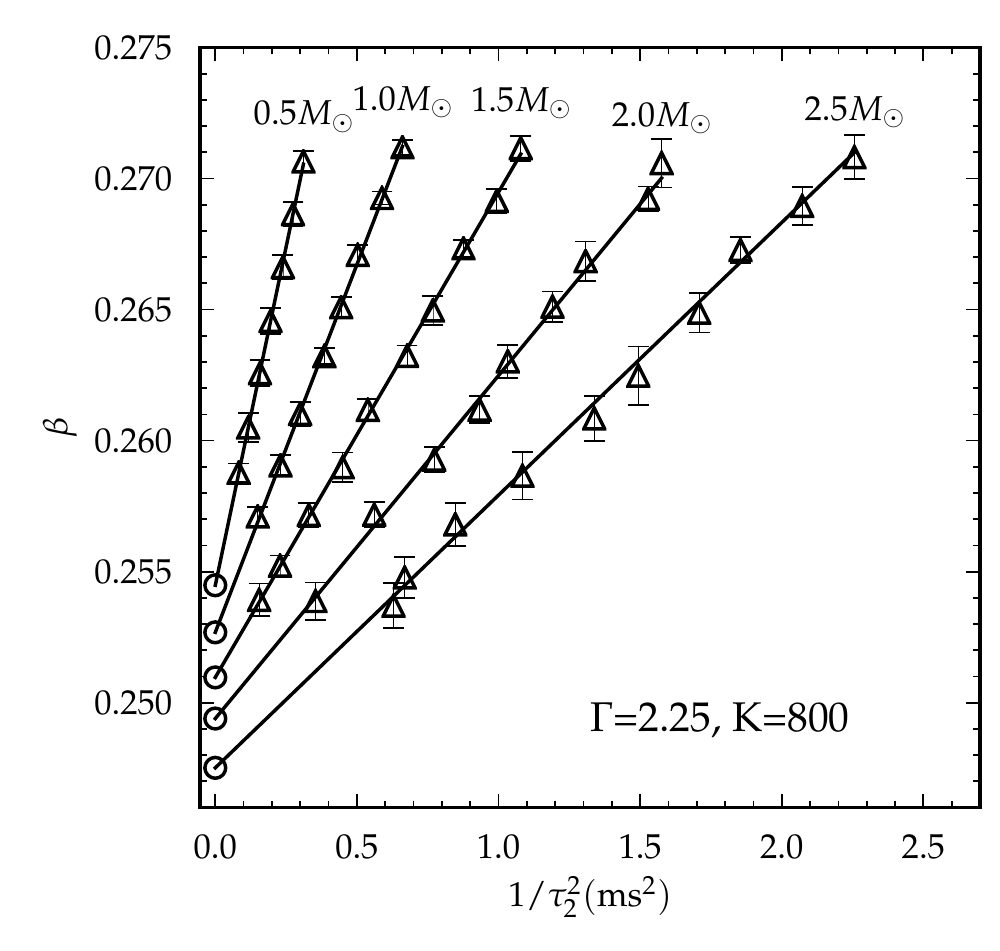} \\[-4mm]
\includegraphics[width=0.45\textwidth]{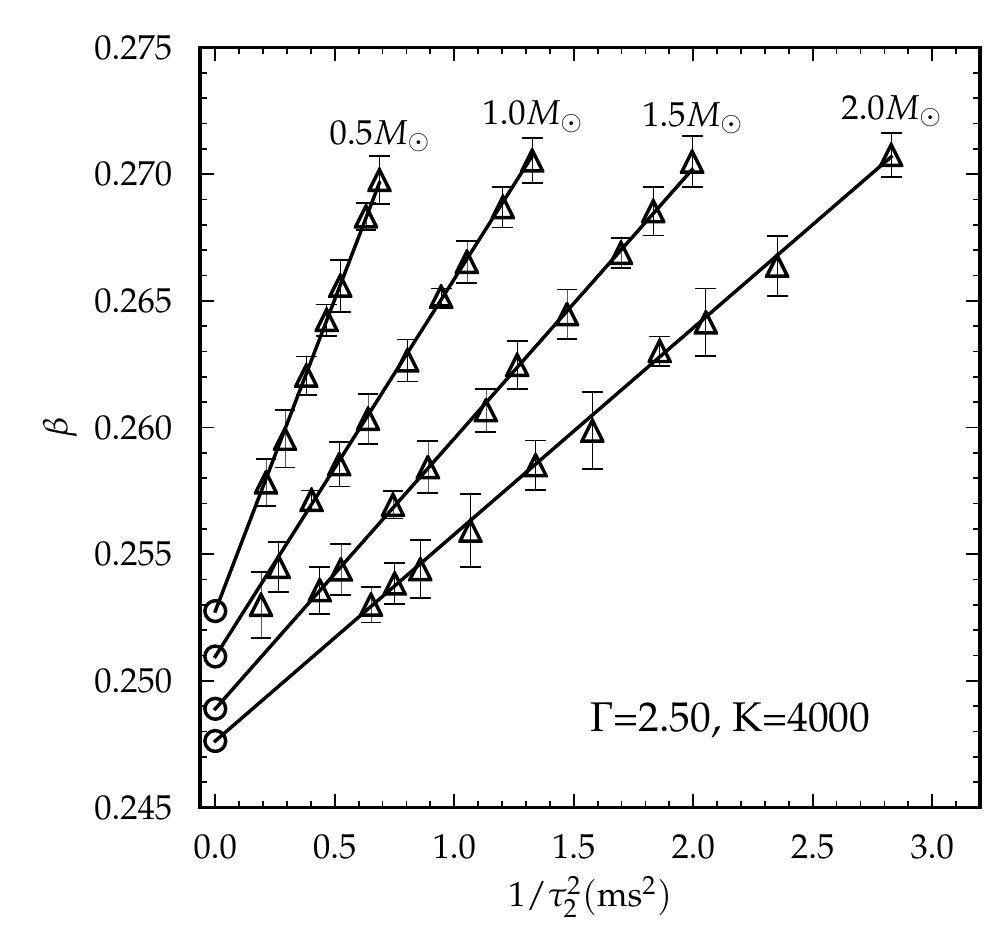}
\includegraphics[width=0.45\textwidth]{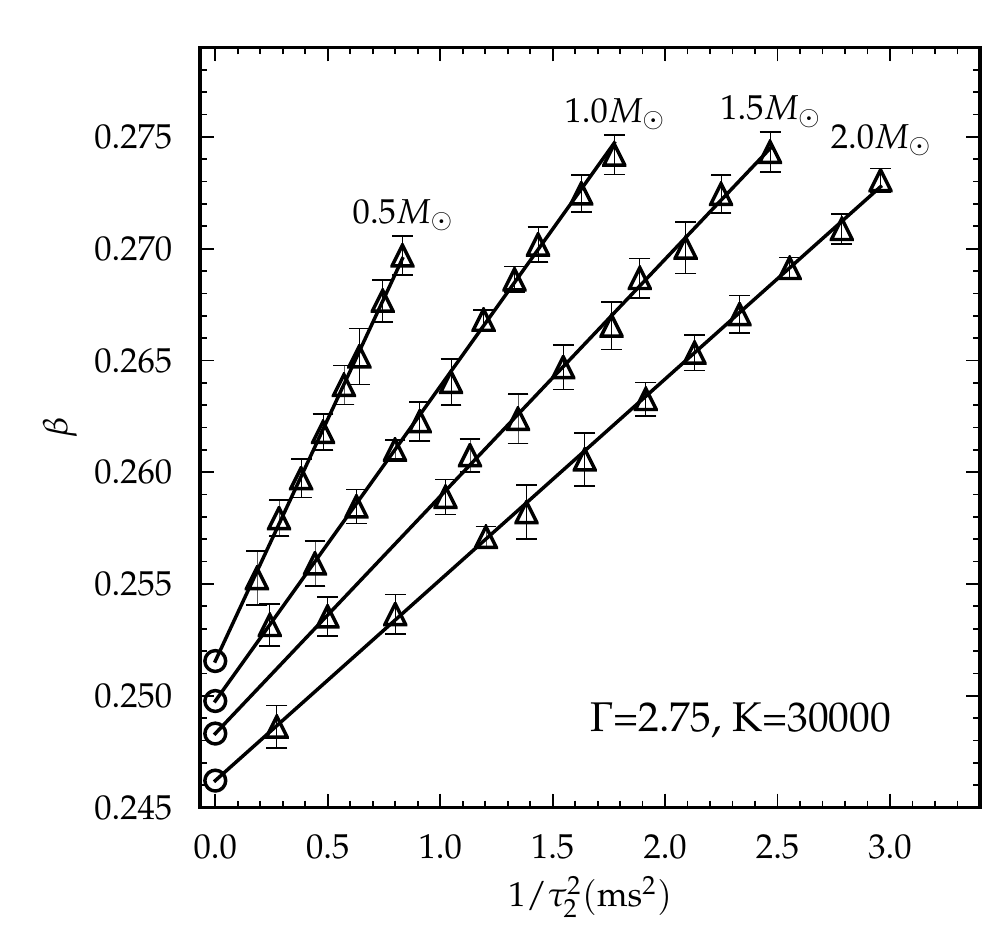} \\[-4mm]
\includegraphics[width=0.45\textwidth]{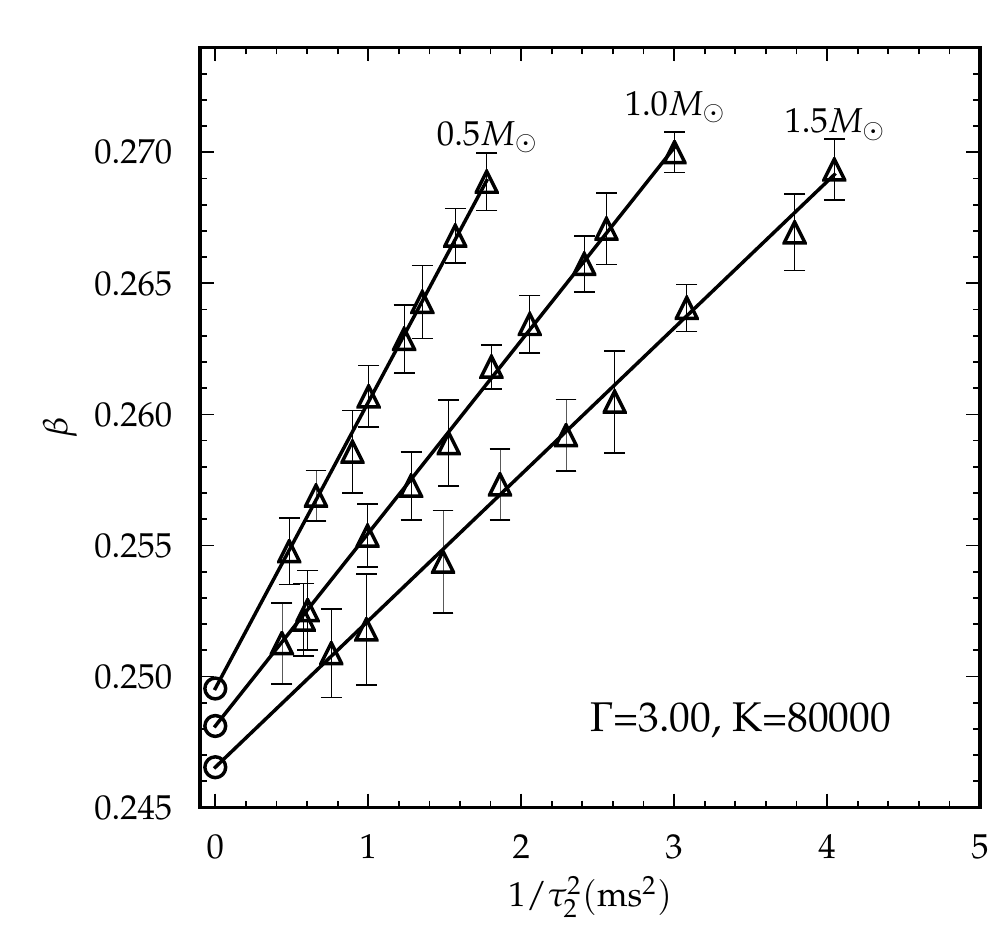}
\includegraphics[width=0.45\textwidth]{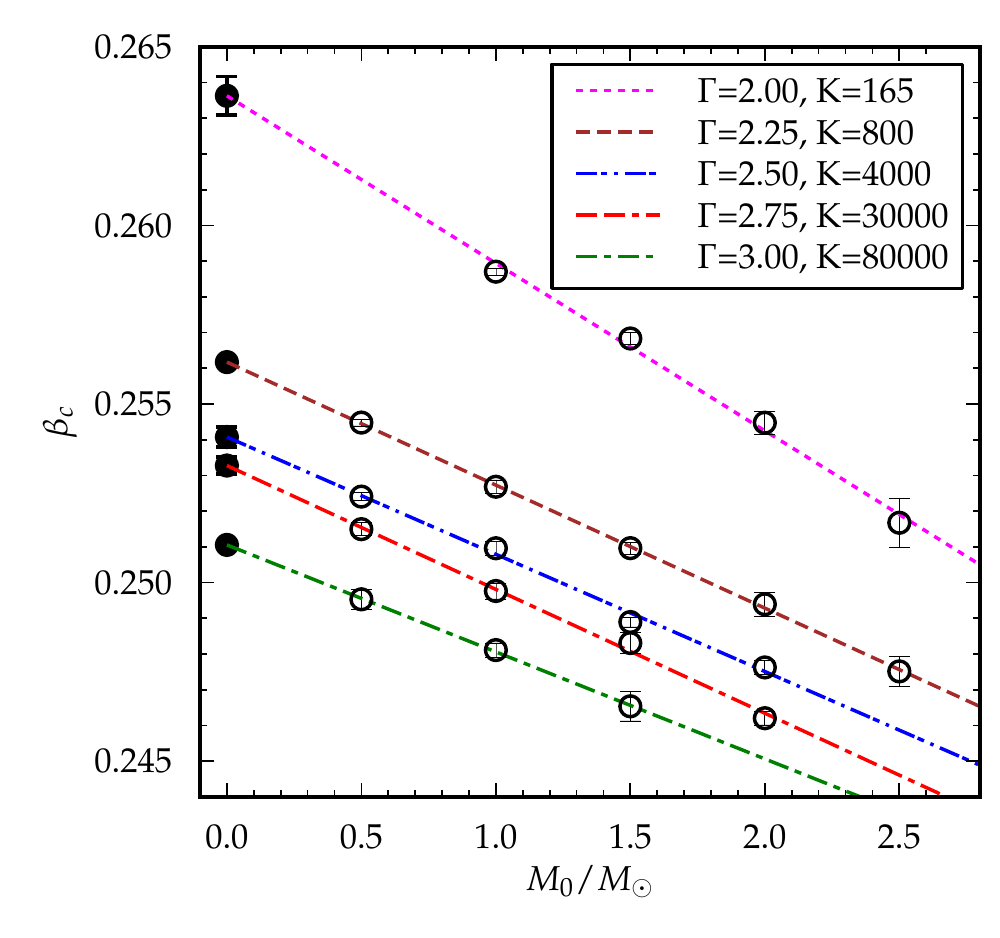}
\end{centering}
\vspace{-4mm}
 \caption{For each value of the adiabatic index $\Gamma$:
   the critical diagram relating the growth time $\tau_2$ of each
   unstable model to the value of the instability parameter
   $\beta$. Triangles represent the values corresponding to all the
   models listed in Tab.~\ref{tab:models}.
   Specifically, we show the dependency of $\beta$ at the beginning of the
   time interval chosen for performing the fit of the $m=2$ mode growth
   (reported in Tab.~\ref{tab:models} as $\beta(t_i)$), on
   $\sfrac{1}{\tau_2^2}$, in order to highlight the quality of the fit.
   For all the
   constant rest-mass sequences considered, we also report, with
   open circles, the extrapolated values $\beta_c$, which are used in
   the bottom right plot to show their dependency on the constant rest-mass,
   as well as the linear fit to the zero mass limit
   using Eq.~(\ref{eq:NewtonianExtrapolation}).
\label{fig:growthtimes}}
\end{figure*}

\begin{figure}
\begin{centering}
\includegraphics[width=0.45\textwidth]{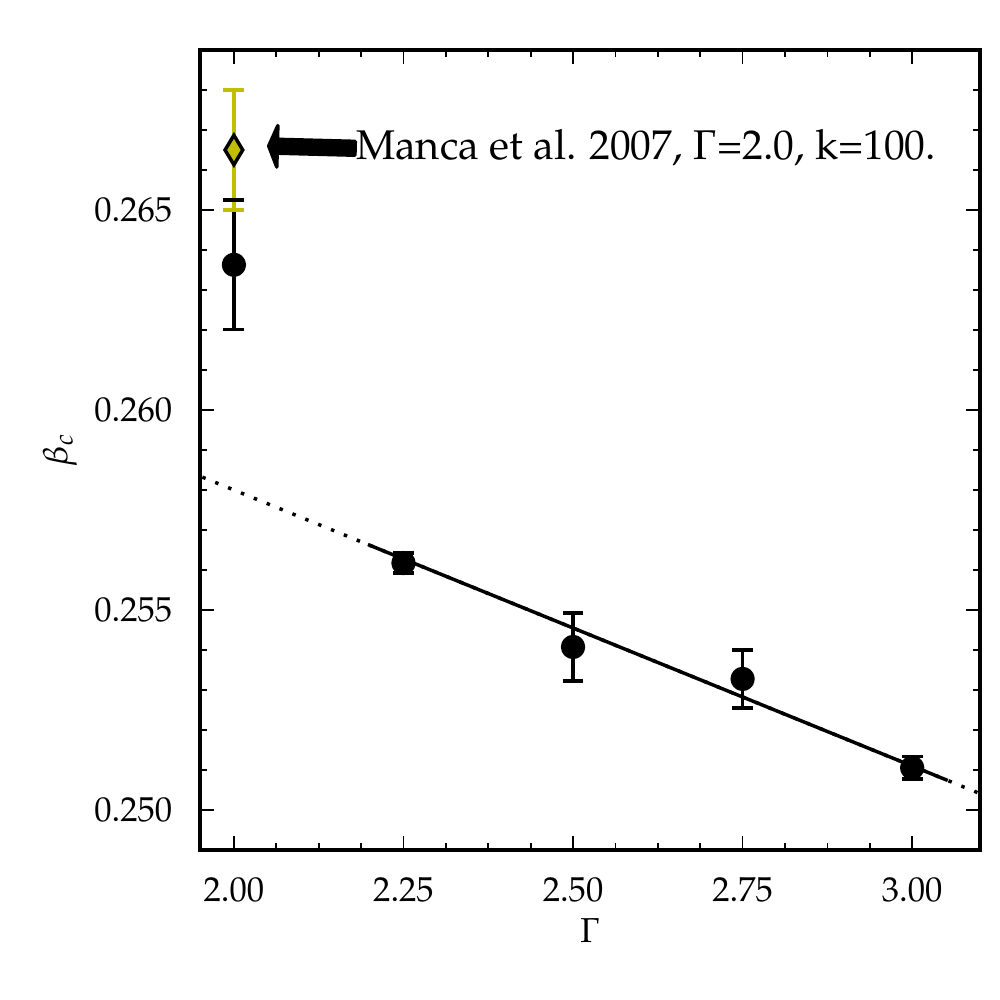}
\end{centering}
 \caption{Dependency of the threshold $\beta_c^N$ of the bar-mode instability
          in the Newtonian limit of zero rest-mass
          on the stiffness of the EOS, i.e. on $\Gamma$, the
          exponent in the polytropic EOS. An approximate
          linear trend is visible in the range $2.25 \lesssim \Gamma\lesssim 3.0$.
          The case of $\Gamma=2.0$ does not follow this trend, however, for reasons
          mentioned in Sec.~\ref{subsec:BAR}. Also mentioned is a result
          from a similar simulation~\cite{Manca:2007ca}, confirming the high value of
          $\beta_c$ for $\Gamma=2.0$.
\label{fig:NewtonianLimit_vs_Gamma}}
\end{figure}

We have performed a fit for the growth time $\tau_2$ of the bar mode
as a function of the instability parameter $\beta$ for twenty-one
sequences of models with constant rest-mass ranging from 0.5 $M_\odot$
to 2.5 $M_\odot$, as shown in Fig.~\ref{fig:growthtimes}. We
estimate the threshold for the onset of the instability using the
extrapolation technique used in~\cite{Manca:2007ca,DePietri:2014mea} where we assume,
in analogy with what expected in the Newtonian case, that the
dependence of the frequency of the mode on $\beta$ is of the type
\begin{equation}
 \sigma(\beta) = \Omega(\beta) \pm \sqrt{F(\beta)}\,,
\end{equation} 
where
\begin{equation}
 F(\beta) = \frac{-1}{\left(\tau_2(\beta)\right)^2} \simeq A (\beta_c-\beta)\, .
\label{eq:res_fits}
\end{equation}

\begin{table}
\begin{tabular}{|c|c|cr@{.}l|cl|}
\hline $\Gamma$ & M$[M_\odot]$ & \multicolumn{3}{c|}{$\beta_c$} & \multicolumn{2}{c|}{$A$} \\

\hline
2.00 & 1.0 && 0&25871(9) && 11.3(1)\\
2.00 & 1.5 && 0&2568(2) && 19.0(3)\\
2.00 & 2.0 && 0&2545(3) && 27.8(8)\\
2.00 & 2.5 && 0&2517(7) && 36(2)\\
\hline
2.25 & 0.5 && 0&25448(9) && 19.4(2)\\
2.25 & 1.0 && 0&2527(2) && 35.6(5)\\
2.25 & 1.5 && 0&2510(2) && 54.0(7)\\
2.25 & 2.0 && 0&2494(3) && 76(2)\\
2.25 & 2.5 && 0&2475(4) && 96(2)\\
\hline
2.50 & 0.5 && 0&2524(1) && 39.6(3)\\
2.50 & 1.0 && 0&2510(2) && 67(1)\\
2.50 & 1.5 && 0&2489(1) && 94(1)\\
2.50 & 2.0 && 0&2476(2) && 123(2)\\
\hline
2.75 & 0.5 && 0&2515(2) && 46.0(7)\\
2.75 & 1.0 && 0&2498(2) && 71(1)\\
2.75 & 1.5 && 0&2483(3) && 94(2)\\
2.75 & 2.0 && 0&2462(2) && 111(1)\\
\hline
3.00 & 0.5 && 0&2495(3) && 92(2)\\
3.00 & 1.0 && 0&2481(2) && 136(2)\\
3.00 & 1.5 && 0&2465(4) && 179(5)\\

 \hline
\end{tabular}
 \caption{Results for the fits of the critical value $\beta_c$ and 
          the slope $A$ (see Eq.~\ref{eq:res_fits}),
          with respect to the total baryon mass $M_0$, and for five seqences of
          various values of the adiabatic EOS index $\Gamma$.
 \label{tab:growthtimes}}
\end{table}

\begin{table}
\begin{tabular}{|c|r@{.}l|r@{.}l|}
\hline $\Gamma$ & \multicolumn{2}{c|}{$\beta_c^{N}$} & \multicolumn{2}{c|}{$q$} \\

\hline
2.00 & 0&2636(5) & 0&0047(3)\\
2.25 & 0&25617(8) & 0&00345(5)\\
2.50 & 0&2541(3) & 0&0033(2)\\
2.75 & 0&2533(2) & 0&0035(2)\\
3.00 & 0&25106(9) & 0&00300(9)\\

 \hline
\end{tabular}
\caption{Results of the fits of the critical value $\beta_c^N$
         ($\beta_c$ in the Newtonian limit of zero baryon mass),
         as well as $q$ (see Eq.~\ref{eq:NewtonianExtrapolation})
         with respect to five different values of the adiabatic index
         $\Gamma$.}
\label{tab:betac}
\end{table}

Results using different polytropic exponents $\Gamma$ cannot be directly compared
to each other to infer the effects of considering a
stiffer EOS. The issue is that when considering a polytropic EOS,
one can change the units of measurement in such a way that the value
of the polytropic constant $K$ is $1$. This means that by changing this
value one effectively changes the mass scale and, in turn, the mass of
the stellar model considered . Indeed, the assertion that for a star
with mass $M_0= 1.0 M_\odot$ the threshold for the onset of the
bar-mode instability is reduced to $0.2498(2)$ for $\Gamma=2.75$ from
the higher value of $0.25871(9)$ for $\Gamma=2.0$ is susceptible to the
choice of the mass scale determined by the choice of the values of the
polytropic constants. The dependence on the choice of the mass
scale can be eliminated by going to the zero-mass limit that
corresponds to performing an extrapolation to the {\it Newtonian} limit
of the results. This can be achieved by a linear fit of the
reported values for the critical $\beta_c$ for the onset of the
classical bar-mode instability in Tab.~(\ref{tab:growthtimes}) as a function
of the baryonic rest-mass (see Fig.~\ref{fig:growthtimes}). The
result for this fit leads to the following expression for the
critical $\beta_c$ as a function of the the total baryonic mass $M_0$:
\begin{align}
  \beta_c(M) & = \beta_c^{N} - q \cdot M_0 \, ,
\label{eq:NewtonianExtrapolation}
\end{align}
with different values of the constant depending on the 
adiabatic index $\Gamma$. These values 
are reported in Tab.~\ref{tab:betac} and shown on
the bottom right box of Figure \ref{fig:growthtimes}.

The extrapolated values for $\beta_c$ in the
limit of zero baryonic mass for the relativistic stellar models then
lead to a dependency on the compactness of the star alone, expressed as
dependency on $\Gamma$, shown in Figure~\ref{fig:NewtonianLimit_vs_Gamma}.
As can be seen there, the dependency of $\beta_c$ on $\Gamma$ is,
within errors, linear in the range $[2.25,3.0]$, while lower values of
$\Gamma$ deviate notably. We also show results from~\cite{Manca:2007ca},
using $\Gamma=2.0$ (and $K=100$), which show a similar deviation. The
fact that the case of $\Gamma=2.0$ is special is not a surprise since 
in the Newtonian limit, i.e., for small values of central density
the equilibrium configuration (see \cite{Chandreaskahr39a}) of a non-rotational
polytrope are described by the Lane-Emden equation, and
the radius of the Star $R$ and its total mass $M$ are related to the central 
density $\rho_c$ as $R\propto\rho_c^{(\Gamma-2)/2}$ and $M\propto\rho_c^{(3 \Gamma-4)/2}$.
That means that the two values $\Gamma=4/3$ and $2$ are very special and represent
the transition points to different behavior of the properties of the associated 
stellar models. In fact, for $\Gamma < \sfrac{4}{3}$ we see that the mass decreases for
increasing central density, and the models can not be stable, while 
$\Gamma = 2$ marks a transition point in the relation between the radius of the star 
and the central density.

Our results show that the dependency of the threshold for the onset of
the dynamical bar-mode instability on $\Gamma$ is not as large as the previously
published results for $\Gamma=2.0$~\cite{Manca:2007ca} and
$\Gamma=2.75$~\cite{DePietri:2014mea} alone suggested, at least not
close to the interesting value of $\Gamma=2.75$. Further investigation
is necessary to clarify the exact dependency at values of $\Gamma$ lower
than $2.25$.

\section{Conclusions}
\label{sec:conclusions}
We have presented a study of the dynamical bar-mode instability in
differentially rotating NSs in full General Relativity for a wide and
systematic range of values of the rotational parameter $\beta$ and the
conserved baryonic mass $M_0$, using a polytropic/ideal-fluid EOS
characterized by a range of values of the adiabatic index $\Gamma=[2.0, 2.25, 2.5, 2.75$, and
$3.0]$.
In particular, we have evolved a large number of NS models belonging
to twenty-one different sequences with a constant rest-mass
ranging from $0.5$ to $2.5 \, M_\odot$, with a fixed degree of
differential rotation ($\hat{A} = 1$), and with many different values
of $\beta$ in the range $[0.255,0.278]$.

For all the models with a sufficiently high initial value of $\beta$
we observe the expected exponential growth of the $m=2$ mode which is
characteristic of the development of the dynamical bar-mode
instability.  We compute the growth time $\tau_2$ for each of these
bar-mode unstable models by performing a nonlinear least-square fit
using a trial function for the quadrupole moment of the matter
distribution.  The growth time clearly depends on both the rest-mass
and the rotation and in particular we find, in agreement with previous
studies~\cite{Manca:2007ca,DePietri:2014mea},
that the relation between
the instability parameter $\beta$ and the inverse square of $\tau_2$,
for each sequence of constant rest-mass, is linear.  

This allows us to extrapolate the threshold value, $\beta_c$, for each sequence
corresponding to the growth time going to infinity, using the same
procedure already employed in \cite{Manca:2007ca,DePietri:2014mea}.
Once the five values of $\beta_c$ for each value of $\Gamma$
have been computed, we are able to show that the dependency of $\beta_c$ on
$\Gamma$ is, within errors, close to linear (see Fig.~\ref{fig:NewtonianLimit_vs_Gamma})
in the the range $2.25 \lesssim \Gamma \lesssim 3.0$.
From this, we
are able to perform a fit, predicting $\beta_c$ for a given value of
$\Gamma$ between $2.25$ and $3$:
\begin{equation}
 \beta_c \approx 0.2718 - 0.0069 \cdot \Gamma \,.
\end{equation}
However, we would like to stress that this should only be taken as rough
estimate. In particular we do not claim an actual linear dependency of
$\beta_c$ on $\Gamma$ in this range. Very likely the dependency is more complex.
Further investigations are needed to clarify the full dependency
of $\beta_c$ on $\Gamma$, especially for values of $\Gamma$ between $2.0$ and
$2.25$.

\acknowledgments

We especially thank N.~Stergioulas for providing us the RNS code that we
used to generate the initial stellar configurations. We are grateful to
Dennis Castleberry, Peter Diener and Steven R.\ Brandt for a careful reading of the manuscript.
We would also like to thank R.~Alfieri, S.~Bernuzzi, N.~Bucciantini, A.~Nagar, L.~Del~Zanna, 
for useful discussions and insights in the development of the present
work. Portions of this research were conducted with high performance computing (HPC)
resources provided by the European Union PRACE program (6$^{th}$ call, project ``3DMagRoI''),
by the Louisiana State University (allocations hpc\_cactus, hpc\_numrel and hpc\_hyrel), 
by the Louisiana Optical Network Initiative (allocations loni\_cactus and loni\_numrel);
by the National Science Foundation through XSEDE resources (allocations TG-ASC120003, 
TG-PHY100033 and TG-MCA02N014), by the INFN ``Theophys'' cluster and through the 
allocation of CPU time on the BlueGene/Q-Fermi at CINECA under the agreement between 
INFN and CINECA. The work of A.~F. has been supported by MIUR (Italy) through the INFN-SUMA project. 
F.~L. is directly supported by, and this project heavily used infrastructure developed using support
from the National Science Foundation in the USA (1212401 / 1212426 / 1212433 / 1212460).
Partial support from INFN ``Iniziativa Specifica TEONGRAV'' and by the “NewCompStar”, COST Action MP1304,
are kindly acknowledged.

\appendix*
\section{Model data}

The results of all fits mentioned in Sec.\ref{subsec:above} are collected in
the following Tab.~\ref{tab:models}, where we report for each model the 
maximum value assumed by the distortion parameter $\max(\eta)$, the time
interval $[t_i,t_f]$ selected for the fit, the value
$\beta(t_i)$ corresponding to the value of the instability parameter
$\beta$ at the beginning of the fit interval and $\tau_2$ and $f_2$,
the growth time and frequency that characterize the $m=2$ bar-mode
instability, respectively.

\begin{longtable}{|c|c|cc|ccc|}
\hline model & $\max(\eta)$ & $t_i$ & $t_f$ & $\beta(t_i)$ & $\tau_2$(ms) & $f_2$(kHz)\\
\hline
\endhead

\hline
G2.00M1.0b0.255 & 0.012 &   -   &   -   &   -     &   -    &   -   \\
G2.00M1.0b0.260 & 0.028 &   -   &   -   &   -     &   -    &   -   \\
G2.00M1.0b0.262 & 0.107 &   -   &   -   &   -     &   -    &   -   \\
G2.00M1.0b0.264 & 0.531 & 26.1 & 43.3 & 0.2633 & 4.388 & 0.251\\
G2.00M1.0b0.266 & 0.949 & 26.5 & 40.1 & 0.2653 & 3.680 & 0.248\\
G2.00M1.0b0.268 & 1.062 & 26.7 & 37.4 & 0.2673 & 3.200 & 0.246\\
G2.00M1.0b0.270 & 1.150 & 24.7 & 35.0 & 0.2694 & 2.879 & 0.244\\
G2.00M1.0b0.272 & 1.226 & 25.3 & 34.7 & 0.2714 & 2.656 & 0.241\\
\hline
G2.00M1.5b0.255 & 0.018 &   -   &   -   &   -     &   -    &   -   \\
G2.00M1.5b0.260 & 0.692 & 22.5 & 40.2 & 0.2596 & 4.421 & 0.342\\
G2.00M1.5b0.262 & 0.860 & 21.7 & 32.4 & 0.2616 & 3.301 & 0.339\\
G2.00M1.5b0.264 & 0.982 & 24.2 & 34.1 & 0.2635 & 2.751 & 0.335\\
G2.00M1.5b0.266 & 1.086 & 21.5 & 29.2 & 0.2655 & 2.460 & 0.332\\
G2.00M1.5b0.268 & 1.171 & 19.1 & 26.9 & 0.2677 & 2.215 & 0.327\\
G2.00M1.5b0.270 & 1.243 & 20.4 & 27.1 & 0.2697 & 2.031 & 0.324\\
G2.00M1.5b0.272 & 1.315 & 19.1 & 25.7 & 0.2716 & 1.867 & 0.321\\
\hline
G2.00M2.0b0.255 & 0.167 &   -   &   -   &   -     &   -    &  -   \\
G2.00M2.0b0.260 & 0.878 & 19.2 & 27.1 & 0.2596 & 2.614 & 0.436\\
G2.00M2.0b0.262 & 0.995 & 15.8 & 24.0 & 0.2617 & 2.239 & 0.430\\
G2.00M2.0b0.264 & 1.086 & 18.8 & 25.3 & 0.2637 & 1.933 & 0.427\\
G2.00M2.0b0.266 & 1.175 & 16.8 & 22.8 & 0.2658 & 1.804 & 0.421\\
G2.00M2.0b0.268 & 1.236 & 16.5 & 22.4 & 0.2677 & 1.637 & 0.413\\
G2.00M2.0b0.270 & 1.306 & 17.5 & 22.8 & 0.2696 & 1.533 & 0.407\\
G2.00M2.0b0.272 & 1.365 & 15.0 & 19.9 & 0.2718 & 1.443 & 0.402\\
\hline
G2.00M2.5b0.260 & 0.977 & 14.8 & 21.5 & 0.2598 & 1.829 & 0.539\\
G2.00M2.5b0.262 & 1.054 & 13.7 & 20.1 & 0.2617 & 1.702 & 0.537\\
G2.00M2.5b0.264 & 1.139 & 15.3 & 20.6 & 0.2637 & 1.488 & 0.523\\
G2.00M2.5b0.266 & 1.219 & 16.8 & 21.8 & 0.2655 & 1.354 & 0.519\\
G2.00M2.5b0.268 & 1.284 & 12.4 & 16.7 & 0.2677 & 1.290 & 0.510\\
G2.00M2.5b0.270 & 1.343 & 13.1 & 17.2 & 0.2697 & 1.257 & 0.500\\
G2.00M2.5b0.272 & 1.384 & 13.7 & 17.7 & 0.2717 & 1.173 & 0.493\\
\hline
\hline
G2.25M0.5b0.255 & 0.006 &   -   &   -   &   -     &   -    &   -   \\
G2.25M0.5b0.260 & 0.492 & 19.6 & 31.5 & 0.2588 & 3.483 & 0.328\\
G2.25M0.5b0.262 & 0.571 & 24.5 & 34.8 & 0.2605 & 2.937 & 0.325\\
G2.25M0.5b0.264 & 0.650 & 19.1 & 27.7 & 0.2626 & 2.519 & 0.323\\
G2.25M0.5b0.266 & 0.717 & 18.9 & 26.5 & 0.2646 & 2.263 & 0.320\\
G2.25M0.5b0.268 & 0.786 & 19.4 & 26.3 & 0.2666 & 2.047 & 0.317\\
G2.25M0.5b0.270 & 0.851 & 16.8 & 23.3 & 0.2686 & 1.910 & 0.314\\
G2.25M0.5b0.272 & 0.909 & 17.9 & 24.0 & 0.2706 & 1.793 & 0.312\\
\hline
G2.25M1.0b0.255 & 0.131 &   -   &   -   &   -     &   -     &   -   \\
G2.25M1.0b0.258 & 0.569 & 16.7 & 25.6 & 0.2571 & 2.585 & 0.470\\
G2.25M1.0b0.260 & 0.747 & 15.4 & 22.3 & 0.2591 & 2.085 & 0.466\\
G2.25M1.0b0.262 & 0.839 & 14.4 & 20.4 & 0.2610 & 1.827 & 0.462\\
G2.25M1.0b0.264 & 0.919 & 12.0 & 17.5 & 0.2632 & 1.611 & 0.458\\
G2.25M1.0b0.266 & 0.983 & 13.8 & 18.9 & 0.2651 & 1.500 & 0.453\\
G2.25M1.0b0.268 & 1.050 & 13.5 & 18.3 & 0.2671 & 1.410 & 0.450\\
G2.25M1.0b0.270 & 1.107 & 12.0 & 16.4 & 0.2693 & 1.303 & 0.444\\
G2.25M1.0b0.272 & 1.165 & 12.7 & 16.9 & 0.2712 & 1.230 & 0.438\\
\hline
G2.25M1.5b0.255 & 0.546 & 14.0 & 21.9 & 0.2539 & 2.542 & 0.612\\
G2.25M1.5b0.256 & 0.658 & 11.7 & 18.3 & 0.2552 & 2.092 & 0.611\\
G2.25M1.5b0.258 & 0.798 & 12.4 & 18.1 & 0.2572 & 1.740 & 0.601\\
G2.25M1.5b0.260 & 0.860 & 13.4 & 18.7 & 0.2590 & 1.491 & 0.597\\
G2.25M1.5b0.262 & 0.970 & 11.9 & 16.4 & 0.2612 & 1.362 & 0.590\\
G2.25M1.5b0.264 & 1.043 & 10.5 & 14.6 & 0.2632 & 1.215 & 0.585\\
G2.25M1.5b0.266 & 1.103 & 12.8 & 16.7 & 0.2650 & 1.141 & 0.578\\
G2.25M1.5b0.268 & 1.170 & 10.1 & 13.7 & 0.2673 & 1.068 & 0.573\\
G2.25M1.5b0.270 & 1.217 & 9.8 & 13.2 & 0.2691 & 1.003 & 0.564\\
G2.25M1.5b0.272 & 1.270 & 9.9 & 13.2 & 0.2712 & 0.963 & 0.557\\
\hline
G2.25M2.0b0.255 & 0.717 & 12.5 & 17.6 & 0.2539 & 1.681 & 0.764\\
G2.25M2.0b0.256 & 0.766 & 11.7 & 16.2 & 0.2549 & 1.536 & 0.754\\
G2.25M2.0b0.258 & 0.892 & 10.6 & 15.1 & 0.2572 & 1.335 & 0.747\\
G2.25M2.0b0.260 & 0.981 & 8.3 & 12.2 & 0.2593 & 1.137 & 0.736\\
G2.25M2.0b0.262 & 1.050 & 8.9 & 12.4 & 0.2612 & 1.035 & 0.731\\
G2.25M2.0b0.264 & 1.109 & 10.2 & 13.7 & 0.2630 & 0.984 & 0.718\\
G2.25M2.0b0.266 & 1.163 & 9.0 & 12.1 & 0.2651 & 0.916 & 0.711\\
G2.25M2.0b0.268 & 1.217 & 9.9 & 12.9 & 0.2668 & 0.875 & 0.702\\
G2.25M2.0b0.270 & 1.266 & 8.1 & 10.8 & 0.2692 & 0.809 & 0.691\\
G2.25M2.0b0.272 & 1.300 & 9.4 & 12.2 & 0.2706 & 0.797 & 0.679\\
\hline
G2.25M2.5b0.254 & 0.698 & 8.2 & 12.6 & 0.2529 & 1.493 & 0.925\\
G2.25M2.5b0.255 & 0.768 & 8.4 & 12.5 & 0.2537 & 1.260 & 0.927\\
G2.25M2.5b0.256 & 0.820 & 8.3 & 12.3 & 0.2548 & 1.223 & 0.918\\
G2.25M2.5b0.258 & 0.907 & 8.8 & 12.5 & 0.2568 & 1.086 & 0.910\\
G2.25M2.5b0.260 & 0.985 & 8.5 & 11.9 & 0.2587 & 0.960 & 0.898\\
G2.25M2.5b0.262 & 1.069 & 7.7 & 10.5 & 0.2608 & 0.864 & 0.878\\
G2.25M2.5b0.264 & 1.114 & 8.3 & 11.1 & 0.2625 & 0.818 & 0.873\\
G2.25M2.5b0.266 & 1.175 & 7.4 & 10.0 & 0.2649 & 0.765 & 0.861\\
G2.25M2.5b0.268 & 1.241 & 7.4 & 9.9 & 0.2673 & 0.734 & 0.842\\
G2.25M2.5b0.270 & 1.276 & 7.5 & 9.8 & 0.2690 & 0.695 & 0.830\\
G2.25M2.5b0.272 & 1.326 & 8.1 & 10.4 & 0.2708 & 0.666 & 0.814\\
\hline
\hline
G2.50M0.5b0.255 & 0.017 &   -   &   -   &   -     &   -    &   -   \\
G2.50M0.5b0.258 & 0.188 &   -   &   -   &   -     &   -    &   -   \\
G2.50M0.5b0.260 & 0.417 & 13.0 & 20.4 & 0.2578 & 2.171 & 0.475\\
G2.50M0.5b0.262 & 0.485 & 14.0 & 20.2 & 0.2596 & 1.852 & 0.471\\
G2.50M0.5b0.264 & 0.541 & 12.1 & 17.7 & 0.2620 & 1.621 & 0.467\\
G2.50M0.5b0.266 & 0.596 & 10.7 & 15.7 & 0.2642 & 1.465 & 0.465\\
G2.50M0.5b0.268 & 0.639 & 13.8 & 18.5 & 0.2656 & 1.382 & 0.460\\
G2.50M0.5b0.270 & 0.697 & 10.8 & 15.1 & 0.2683 & 1.259 & 0.456\\
G2.50M0.5b0.272 & 0.744 & 12.3 & 16.4 & 0.2698 & 1.207 & 0.451\\
\hline
G2.50M1.0b0.255 & 0.449 & 12.6 & 20.4 & 0.2530 & 2.288 & 0.656\\
G2.50M1.0b0.256 & 0.507 & 9.9 & 16.6 & 0.2545 & 1.942 & 0.654\\
G2.50M1.0b0.258 & 0.618 & 9.6 & 15.0 & 0.2571 & 1.575 & 0.650\\
G2.50M1.0b0.260 & 0.695 & 10.0 & 14.6 & 0.2585 & 1.389 & 0.644\\
G2.50M1.0b0.262 & 0.760 & 10.7 & 14.9 & 0.2603 & 1.251 & 0.637\\
G2.50M1.0b0.264 & 0.838 & 8.9 & 12.7 & 0.2626 & 1.116 & 0.632\\
G2.50M1.0b0.266 & 0.898 & 8.6 & 12.1 & 0.2652 & 1.029 & 0.627\\
G2.50M1.0b0.268 & 0.944 & 9.7 & 13.0 & 0.2665 & 0.975 & 0.621\\
G2.50M1.0b0.270 & 1.007 & 7.8 & 11.0 & 0.2687 & 0.912 & 0.615\\
G2.50M1.0b0.272 & 1.062 & 8.3 & 11.3 & 0.2705 & 0.868 & 0.608\\
\hline
G2.50M1.5b0.255 & 0.642 & 8.0 & 12.9 & 0.2536 & 1.513 & 0.829\\
G2.50M1.5b0.256 & 0.682 & 9.1 & 13.6 & 0.2544 & 1.380 & 0.817\\
G2.50M1.5b0.258 & 0.769 & 8.8 & 12.9 & 0.2569 & 1.160 & 0.812\\
G2.50M1.5b0.260 & 0.841 & 8.2 & 11.8 & 0.2584 & 1.060 & 0.806\\
G2.50M1.5b0.262 & 0.920 & 7.1 & 10.3 & 0.2607 & 0.940 & 0.796\\
G2.50M1.5b0.264 & 0.977 & 8.1 & 11.1 & 0.2625 & 0.890 & 0.790\\
G2.50M1.5b0.266 & 1.031 & 8.2 & 11.0 & 0.2645 & 0.824 & 0.781\\
G2.50M1.5b0.268 & 1.080 & 8.1 & 10.7 & 0.2669 & 0.767 & 0.770\\
G2.50M1.5b0.270 & 1.131 & 7.1 & 9.6 & 0.2685 & 0.739 & 0.763\\
G2.50M1.5b0.272 & 1.184 & 6.2 & 8.7 & 0.2705 & 0.708 & 0.754\\
\hline
G2.50M2.0b0.254 & 0.657 & 6.1 & 10.3 & 0.2530 & 1.239 & 1.011\\
G2.50M2.0b0.255 & 0.732 & 6.5 & 10.5 & 0.2538 & 1.154 & 1.001\\
G2.50M2.0b0.256 & 0.762 & 7.4 & 11.1 & 0.2544 & 1.080 & 0.998\\
G2.50M2.0b0.258 & 0.816 & 7.8 & 11.0 & 0.2559 & 0.968 & 0.985\\
G2.50M2.0b0.260 & 0.910 & 7.3 & 10.2 & 0.2585 & 0.864 & 0.977\\
G2.50M2.0b0.262 & 0.967 & 6.7 & 9.2 & 0.2599 & 0.796 & 0.967\\
G2.50M2.0b0.264 & 1.028 & 6.1 & 8.6 & 0.2630 & 0.733 & 0.953\\
G2.50M2.0b0.266 & 1.089 & 6.6 & 9.0 & 0.2641 & 0.698 & 0.941\\
G2.50M2.0b0.268 & 1.133 & 5.8 & 8.1 & 0.2664 & 0.652 & 0.929\\
G2.50M2.0b0.272 & 1.240 & 5.6 & 7.6 & 0.2708 & 0.595 & 0.901\\
\hline
G2.50M2.5b0.254 & 0.599 & 5.9 & 9.8 & 0.2521 & 1.191 & 1.215\\
G2.50M2.5b0.255 & 0.657 & 6.6 & 10.1 & 0.2529 & 1.106 & 1.214\\
G2.50M2.5b0.256 & 0.715 & 7.2 & 10.6 & 0.2535 & 0.951 & 1.209\\
G2.50M2.5b0.258 & 0.794 & 7.4 & 10.6 & 0.2555 & 0.938 & 1.197\\
G2.50M2.5b0.260 & 0.909 & 5.5 & 7.9 & 0.2584 & 0.750 & 1.176\\
G2.50M2.5b0.262 & 0.973 & 6.5 & 8.8 & 0.2599 & 0.714 & 1.159\\
G2.50M2.5b0.264 & 1.025 & 6.7 & 8.9 & 0.2615 & 0.668 & 1.137\\
G2.50M2.5b0.266 & 1.081 & 5.9 & 8.0 & 0.2641 & 0.622 & 1.124\\
G2.50M2.5b0.268 & 1.142 & 5.5 & 7.5 & 0.2663 & 0.577 & 1.104\\
G2.50M2.5b0.270 & 1.181 & 5.8 & 7.7 & 0.2681 & 0.559 & 1.087\\
G2.50M2.5b0.272 & 1.246 & 5.6 & 7.3 & 0.2703 & 0.522 & 1.060\\
\hline
\hline
G2.75M0.5b0.255 & 0.180 &   -   &   -   &   -     &   -    &   -   \\
G2.75M0.5b0.258 & 0.341 & 14.3 & 22.4 & 0.2553 & 2.325 & 0.517\\
G2.75M0.5b0.260 & 0.415 & 10.7 & 16.9 & 0.2579 & 1.878 & 0.515\\
G2.75M0.5b0.262 & 0.473 & 12.0 & 17.4 & 0.2597 & 1.619 & 0.513\\
G2.75M0.5b0.264 & 0.524 & 10.8 & 15.7 & 0.2618 & 1.445 & 0.509\\
G2.75M0.5b0.266 & 0.578 & 9.2 & 13.7 & 0.2639 & 1.322 & 0.505\\
G2.75M0.5b0.268 & 0.616 & 11.8 & 16.1 & 0.2652 & 1.249 & 0.501\\
G2.75M0.5b0.270 & 0.666 & 9.9 & 13.8 & 0.2677 & 1.159 & 0.498\\
G2.75M0.5b0.272 & 0.707 & 9.9 & 13.7 & 0.2697 & 1.096 & 0.493\\
\hline
G2.75M1.0b0.255 & 0.480 & 10.4 & 17.0 & 0.2532 & 2.031 & 0.685\\
G2.75M1.0b0.258 & 0.618 & 9.9 & 15.0 & 0.2559 & 1.503 & 0.678\\
G2.75M1.0b0.260 & 0.709 & 7.6 & 11.9 & 0.2585 & 1.262 & 0.673\\
G2.75M1.0b0.262 & 0.790 & 6.6 & 10.4 & 0.2610 & 1.119 & 0.669\\
G2.75M1.0b0.264 & 0.830 & 8.6 & 12.2 & 0.2623 & 1.049 & 0.661\\
G2.75M1.0b0.266 & 0.888 & 8.8 & 12.0 & 0.2640 & 0.977 & 0.656\\
G2.75M1.0b0.268 & 0.945 & 7.0 & 10.0 & 0.2668 & 0.916 & 0.654\\
G2.75M1.0b0.270 & 0.996 & 7.2 & 10.2 & 0.2686 & 0.867 & 0.647\\
G2.75M1.0b0.272 & 1.042 & 8.4 & 11.2 & 0.2702 & 0.835 & 0.639\\
G2.75M1.0b0.274 & 1.097 & 6.7 & 9.3 & 0.2725 & 0.784 & 0.633\\
G2.75M1.0b0.276 & 1.131 & 7.4 & 9.9 & 0.2742 & 0.751 & 0.624\\
\hline
G2.75M1.5b0.255 & 0.664 & 7.7 & 12.3 & 0.2535 & 1.416 & 0.839\\
G2.75M1.5b0.260 & 0.867 & 6.9 & 10.3 & 0.2589 & 0.988 & 0.816\\
G2.75M1.5b0.262 & 0.926 & 7.4 & 10.5 & 0.2608 & 0.940 & 0.808\\
G2.75M1.5b0.264 & 0.989 & 6.8 & 9.7 & 0.2624 & 0.862 & 0.801\\
G2.75M1.5b0.266 & 1.034 & 7.1 & 9.8 & 0.2647 & 0.804 & 0.795\\
G2.75M1.5b0.268 & 1.082 & 7.0 & 9.6 & 0.2666 & 0.753 & 0.788\\
G2.75M1.5b0.270 & 1.151 & 6.7 & 9.2 & 0.2687 & 0.728 & 0.778\\
G2.75M1.5b0.272 & 1.163 & 7.2 & 9.5 & 0.2700 & 0.692 & 0.771\\
G2.75M1.5b0.274 & 1.241 & 7.2 & 9.4 & 0.2724 & 0.667 & 0.757\\
G2.75M1.5b0.276 & 1.269 & 6.8 & 8.9 & 0.2743 & 0.637 & 0.746\\
\hline
G2.75M2.0b0.250 & 0.411 & 6.9 & 12.6 & 0.2486 & 1.912 & 1.011\\
G2.75M2.0b0.255 & 0.748 & 6.6 & 10.3 & 0.2536 & 1.118 & 0.988\\
G2.75M2.0b0.258 & 0.876 & 5.7 & 8.8 & 0.2571 & 0.912 & 0.978\\
G2.75M2.0b0.260 & 0.923 & 6.7 & 9.6 & 0.2582 & 0.850 & 0.968\\
G2.75M2.0b0.262 & 0.999 & 6.4 & 9.2 & 0.2606 & 0.780 & 0.956\\
G2.75M2.0b0.264 & 1.059 & 5.9 & 8.3 & 0.2633 & 0.723 & 0.945\\
\newpage
G2.75M2.0b0.266 & 1.118 & 5.2 & 7.5 & 0.2653 & 0.685 & 0.933\\
G2.75M2.0b0.268 & 1.160 & 5.9 & 8.1 & 0.2671 & 0.655 & 0.922\\
G2.75M2.0b0.270 & 1.192 & 5.2 & 7.3 & 0.2691 & 0.626 & 0.912\\
G2.75M2.0b0.272 & 1.249 & 5.5 & 7.5 & 0.2709 & 0.599 & 0.900\\
G2.75M2.0b0.274 & 1.294 & 5.3 & 7.2 & 0.2731 & 0.581 & 0.883\\
\hline
\hline

G3.00M0.5b0.255 & 0.139 &  -   &   -   &   -     &   -    &   -   \\
G3.00M0.5b0.256 & 0.179 &  -   &   -   &   -     &   -    &   -  \\
G3.00M0.5b0.258 & 0.255 & 7.1 & 12.0 & 0.2548 & 1.439 & 0.767\\
G3.00M0.5b0.260 & 0.293 & 7.4 & 11.4 & 0.2569 & 1.232 & 0.763\\
G3.00M0.5b0.262 & 0.345 & 6.0 & 9.5 & 0.2586 & 1.056 & 0.760\\
G3.00M0.5b0.264 & 0.367 & 7.7 & 11.1 & 0.2607 & 0.999 & 0.752\\
G3.00M0.5b0.266 & 0.415 & 6.2 & 9.2 & 0.2629 & 0.900 & 0.747\\
G3.00M0.5b0.268 & 0.442 & 8.0 & 10.8 & 0.2643 & 0.859 & 0.741\\
G3.00M0.5b0.270 & 0.478 & 6.9 & 9.6 & 0.2668 & 0.798 & 0.738\\
G3.00M0.5b0.272 & 0.514 & 5.6 & 8.1 & 0.2689 & 0.751 & 0.731\\
\hline
G3.00M1.0b0.254 & 0.312 & 4.9 & 9.5 & 0.2513 & 1.516 & 1.020\\
G3.00M1.0b0.255 & 0.363 & 6.1 & 10.4 & 0.2522 & 1.314 & 1.012\\
G3.00M1.0b0.256 & 0.363 & 9.3 & 13.4 & 0.2525 & 1.287 & 1.002\\
G3.00M1.0b0.258 & 0.462 & 6.2 & 9.8 & 0.2554 & 1.002 & 0.997\\
G3.00M1.0b0.260 & 0.535 & 6.2 & 9.1 & 0.2573 & 0.884 & 0.995\\
G3.00M1.0b0.262 & 0.590 & 6.2 & 8.9 & 0.2589 & 0.809 & 0.985\\
G3.00M1.0b0.264 & 0.641 & 5.8 & 8.4 & 0.2618 & 0.744 & 0.975\\
G3.00M1.0b0.266 & 0.686 & 6.0 & 8.3 & 0.2634 & 0.697 & 0.967\\
G3.00M1.0b0.268 & 0.738 & 4.7 & 6.8 & 0.2657 & 0.644 & 0.962\\
G3.00M1.0b0.270 & 0.781 & 6.3 & 8.4 & 0.2671 & 0.625 & 0.951\\
G3.00M1.0b0.272 & 0.829 & 5.2 & 7.2 & 0.2700 & 0.577 & 0.941\\
\hline
G3.00M1.5b0.254 & 0.418 & 4.8 & 8.4 & 0.2509 & 1.148 & 1.258\\
G3.00M1.5b0.256 & 0.498 & 7.3 & 10.4 & 0.2518 & 1.006 & 1.237\\
G3.00M1.5b0.258 & 0.574 & 5.6 & 8.5 & 0.2544 & 0.820 & 1.236\\
G3.00M1.5b0.260 & 0.651 & 4.7 & 7.1 & 0.2573 & 0.733 & 1.228\\
G3.00M1.5b0.262 & 0.723 & 5.3 & 7.6 & 0.2592 & 0.660 & 1.211\\
G3.00M1.5b0.264 & 0.755 & 5.2 & 7.3 & 0.2605 & 0.619 & 1.195\\
G3.00M1.5b0.266 & 0.830 & 3.2 & 5.1 & 0.2641 & 0.570 & 1.188\\
G3.00M1.5b0.270 & 0.917 & 5.1 & 6.8 & 0.2669 & 0.514 & 1.156\\
G3.00M1.5b0.272 & 0.967 & 5.1 & 6.8 & 0.2693 & 0.497 & 1.141\\
\hline

\caption{
Results for various quantities for all models, sorted by value of $\Gamma$, mass and $\beta$.
We report for each model the
maximum value assumed by the distortion parameter $\max(\eta)$, the time
interval $[t_i,t_f]$ selected for the fit, the value
$\beta(t_i)$ corresponding to the value of the instability parameter
$\beta$ at the beginning of the fit interval and $\tau_2$ and $f_2$,
the growth time and frequency that characterize the $m=2$ bar-mode
instability, respectively.}
\label{tab:models}
\end{longtable}

\end{document}